\documentclass[12pt,preprint]{aastex}
\usepackage{natbib}
\def\lineunits{ergs\ s$^{-1}$\,cm$^{-2}$}
\def\contunits{ergs\ s$^{-1}$\,cm$^{-2}$\,\AA$^{-1}$}

\def\kms{\ifmmode {\rm km\ s}^{-1} \else km s$^{-1}$\fi}
\def\Hbeta{\ifmmode {\rm H}\beta \else H$\beta$\fi}
\def\Hgamma{\ifmmode {\rm H}\gamma \else H$\gamma$\fi}
\def\heii{He\,{\sc ii}}
\def\hei{He\,{\sc i}}
\def\feii{Fe\,{\sc ii}}

\shorttitle{Variability of Fe\,{\small\bf II} in NGC 5548}
\shortauthors{Vestergaard and Peterson}

\begin{document}
\title{Variability of Fe\,{\small\bf II} Emission Features in the Seyfert 1 
Galaxy NGC 5548}

\author{Marianne~Vestergaard\altaffilmark{1,2} and
Bradley~M.~Peterson\altaffilmark{1}}

\altaffiltext{1}{Department of Astronomy, The Ohio State
University, 140 West 18th Avenue, Columbus, OH  43210}
\altaffiltext{2}{Steward Observatory,
University of Arizona, 933 N. Cherry Avenue, Tucson, AZ 85721}

\email{mvestergaard@as.arizona.edu; peterson@astronomy.ohio-state.edu}

\begin{abstract}
We study the low-contrast Fe\,{\sc ii} emission
blends in the ultraviolet (1250--2200\,\AA) and optical
(4000--6000\,\AA) spectra of the Seyfert 1 galaxy NGC 5548
and show that these features vary in flux and that these
variations are correlated with those of the optical continuum.
The amplitude of variability of the optical Fe\,{\sc ii} emission
is 50\% -- 75\% 
that of \Hbeta\ and the ultraviolet Fe\,{\sc ii} emission
varies with an even larger amplitude than \Hbeta. However, 
accurate measurement
of the flux in these blends proves to be very difficult even
using excellent Fe\,{\sc ii} templates to fit the spectra.
We are able to constrain only weakly
the optical Fe\,{\sc ii} emission-line
response timescale to a value less than several weeks; this
upper limit exceeds all the reliably measured emission-line lags
in this source so it is not particularly meaningful.
Nevertheless, the fact that the optical Fe\,{\sc ii} and continuum flux
variations are correlated
indicates that line fluorescence in a photoionized plasma,
rather than collisional excitation, is responsible for the Fe\,{\sc ii}
emission.
The iron emission templates are available upon request.
\end{abstract}


\section{INTRODUCTION}
Emission from singly ionized iron is one of the principal 
coolants of the broad-line region (BLR) in active galactic 
nuclei (AGNs): indeed, the prominent blend of Fe\,{\sc ii} and 
Balmer continuum emission in the near UV spectrum, a feature 
often referred to as the ``small blue bump,'' produces as much 
as one third of the line flux in some objects 
(Wills, Netzer, \& Wills 1985; Maoz et al.\ 1993). 
Remarkably, however, the characteristics of Fe\,{\sc ii} 
emission in AGNs are very poorly understood from both 
theoretical and observational perspectives:

\begin{enumerate}

\item {\em The excitation mechanism that produces Fe\,{\small\it II} 
emission in AGN spectra is not known.} 
Theoretical models, based on either photoionization or
collisional excitation, have for a long time fallen short in 
reproducing the observed Fe\,{\sc ii} emission (e.g., Collin \& Joly 
2000; Baldwin et al.\ 2004).  Also, conflicting results appear in the 
literature as to the importance of the hard X-rays as an excitation
agent as originally anticipated by Davidson \& Netzer (1979) 
and as to that of the slope of the soft X-ray spectrum 
(e.g., Wilkes, Elvis, \& M$^{\rm c}$Hardy 1987; Shastri et al.\ 1994;
Boroson 1989; Zheng \& O'Brien 1990; see also Wilkes et al.\ 1999).
Even with much-improved current models, either significant 
microturbulence (or velocity shear) in a photoionized medium or a 
contribution from a collisionally ionized gas seems 
to be required to account for the
strength and shape of Fe\,{\sc ii} emission in AGN spectra
(e.g., Baldwin et al.\ 2004).  The conditions that 
are necessary to reproduce the observed Fe\,{\sc ii} spectra 
do not provide clear and straightforward
diagnostics to distinguish between these possibilities.

\item {\em Strong, broad  Fe\,{\small\it II} emission pervades the UV 
through IR spectra of AGNs, contaminating other spectral 
features.}
The complex atomic structure of Fe$^+$ results in 
tens of thousands
of transitions (e.g., Verner et al.\ 1999; Sigut \& Pradhan 2003) 
that produce blended features across the UV, optical, and 
IR spectrum. Isolating individual emission lines under 
these circumstances is challenging, leading to systematic 
uncertainties in measurement of emission-line fluxes, widths, 
and profiles.
The intrinsically large line widths of AGN emission lines
add to the complexity of this issue.

\end{enumerate}
Further underscoring the importance of understanding AGN Fe\,{\sc 
ii} emission is the possibility of using AGN spectra 
to measure the cosmic evolution of the abundance of iron, thus 
providing potentially strong constraints on the history of 
chemical enrichment due to star formation (e.g., Hamann \& 
Ferland 1999; Dietrich et al. 2003 and references therein, but see
also Verner et al.\ 2004; Baldwin et al.\ 2004).

A potentially powerful way of exploring the Fe\,{\sc ii} 
emitting region in AGNs is through flux variability,
specifically by emission-line reverberation mapping (Blandford 
\& McKee 1982; Peterson 1993, 2001). At minimum, 
reverberation mapping can be expected to yield the typical 
distance of the Fe\,{\sc ii}-emitting gas from the central 
continuum source through measurement of the time-delayed 
response of the emission lines to continuum variations. 
This would allow us to constrain  observationally
some of the physical conditions conducive to strong Fe\,{\sc ii}
emission. For example, variability time scales for the optical
and UV Fe\,{\sc ii} emission, respectively, will help discern
whether or not the multiplets at these energies originate in a
common region. This is a key issue for understanding their 
relative intensity ratio in terms of current Fe\,{\sc ii} models
which fail to reproduce observations (Baldwin et al.\ 2004).
At the present time, studies of Fe\,{\sc ii} flux variability 
have been very limited:

\begin{enumerate}
\item Maoz et al. (1993) used UV data obtained with the {\em 
International Ultraviolet Explorer (IUE)} and optical data 
obtained with multiple ground-based telescopes by the 
International AGN Watch\footnote{All International AGN 
Watch data are available at 
http://www.astronomy.ohio-state.edu/$\sim$agnwatch.} 
over an 8-month period in 1989 to study 
the variability of the small blue bump in NGC 5548. They 
concluded that the small blue bump emission arises 
approximately at the same distance as the Ly$\alpha$ emission, 
about 10 light days from the continuum source. While in many 
ways, this represents the best study to date on variability of the 
small blue bump, it must be noted that (a) the quality of {\em 
IUE} data in this part of the spectrum (i.e., 
obtained with the LWP camera) 
is quite poor and (b) there are relatively few epochs in 
which there is no gap in wavelength coverage
between the UV and optical spectra, 
making measurements of the small blue bump variability 
systematically somewhat uncertain.
\item Goad et al.\ (1999) observed the small blue bump in NGC 
3516 on six occasions over an 11-month period in 1995--96. 
Remarkably, while the continuum and high-ionization emission 
lines varied in flux
by a factor of 2, the Mg\,{\sc ii}\,$\lambda2800$ and Fe\,{\sc ii} 
UV emission did not vary by more than 7\% over this time.
\item Kollatschny, Bischoff, \& Dietrich (2000) carried out a 
20-year study of emission-line variations in NGC 7603 and 
found that the optical Fe\,{\sc ii} blends vary with an 
amplitude comparable to that of the Balmer lines.
\item Kollatschny \& Welsh (2001) searched for and found no
optical Fe\,{\sc ii} variability in Mrk 110
to within their sensitivity of 10\%, 
although clear optical Fe\,{\sc ii} variability is 
seen in the case of Fairall 9 
with an amplitude half that of \Hbeta\ (Kollatschny \& Fricke 1985).
\item
Sergeev et al.\ (1997) attempt to constrain the time delay of the
Fe\,{\sc ii} multiplets in the wings of [O\,{\sc iii}]\,$\lambda$5007
in the spectrum of NGC\,5548. They 
find a time delay of a few hundred days. There are, however,
a number of serious difficulties with this result arising from
strong blending of lines, data quality, and 
time-series problems.
\item
A few other studies make claims of Fe\,{\sc ii} 
variability (Kollatschny et al.\ 1981; Salamanca, Alloin, \& Pelat 1995;
Doroshenko et al.\ 1999). On the basis of spectra of
narrow-line Seyfert 1 (NLS1) galaxies taken a year apart,
Giannuzzo \& Stirpe (1996) find that the
optical Fe\,{\sc ii} does seem to vary in these sources.
For the three sources with detected 
variability in both \Hbeta\ and Fe\,{\sc ii}, the Fe\,{\sc ii}/\Hbeta\
flux ratio ranged from 0.25 to 1.0; this would put
Fe\,{\sc ii} variability below the detection threshold for the
sources in which it was not detected.
\end{enumerate}

Given the importance of understanding Fe\,{\sc ii} emission in 
AGNs and given the limited attention this topic has received in 
the past, we have undertaken an examination of Fe\,{\sc ii} 
variability in NGC 5548. In terms of emission-line variability, 
NGC 5548 is by far the best-studied AGN (see Peterson et al.\ 
2002 and references therein). On the other hand, as Baldwin et 
al.\ (2004) point out, Fe\,{\sc ii} emission is not especially 
strong in this source and other sources may be better 
suited to such a study. While we agree with this statement, we 
also feel that the wealth of data on other lines in NGC 5548 
affords a special context in which to examine Fe\,{\sc ii} 
variability and, of course, suitable data for this study already 
exist. Moreover, cursory consideration of Fe\,{\sc ii} 
variability in NGC 5548 yields mixed impressions. For 
example, the difference between spectra taken at recent 
historical maximum and minimum flux states (see Peterson et 
al.\ 1999, Figure 3) seems to clearly show optical Fe\,{\sc ii} 
emission, indicating that these blends do indeed vary in this 
source. However, Kollatschny \& Welsh (2001)
attribute the apparent variations solely to changes
in He\,{\sc ii}\,$\lambda4686$. If indeed the optical Fe\,{\sc ii} 
emission varies with time, then we ought to see evidence of these 
features in the difference or rms spectra (defined below in \S{2.3.2.})
that we use to isolate the variable parts of the emission lines 
(e.g., Peterson et al.\ 2004); our initial impression is that the 
optical Fe\,{\sc ii} blends do not appear in rms spectra and thus 
there is no compelling evidence for optical Fe\,{\sc ii} 
variability, at least on time scales of several months.

Thus, the goal of this study is to determine whether or not 
Fe\,{\sc ii} emission varies in NGC 5548 and, if so, with what 
amplitude and on what timescale. We also attempt to 
constrain meaningfully the Fe\,{\sc ii} line width as an 
additional test of the common assumption that the optical 
Fe\,{\sc ii} emission arises co-spatially with the Balmer lines 
(e.g., Phillips 1978). We will not specifically re-examine the 
issue of the variability of the small blue bump since the data 
examined by Maoz et al.\ (1993) have not been superseded by 
superior data and we are unlikely to be able to improve on their 
results. We will thus concentrate on variability of the optical 
Fe\,{\sc ii} features, 
comprised primarily of two blends, one between
\Hgamma\ and \Hbeta, and the other covering the
spectral ranges 5100--5700\,\AA\
(often referred to as the 4570\,\AA\ and 5190, 5320\,\AA\ blends,
respectively; Osterbrock 1976),
and of the UV Fe\,{\sc ii} blends in the 
vicinity of the C\,{\sc iv}\,$\lambda1549$ emission line.

\section{METHODOLOGY AND DATA ANALYSIS}

\subsection{Data}
As noted above, we are focusing on NGC 5548 on account of 
the large amount of variability data available. There are two 
suitable UV data sets. The first of these is a set of {\em IUE} 
spectra obtained over an 8-month period in 1988--89, sampled 
once every four days (Clavel et al.\ 1991). The second set of 
UV spectra were obtained with the Faint Object Spectrograph 
(FOS) on {\em Hubble Space Telescope (HST)} over a 39-day 
period in 1993, sampled once per day (Korista et al.\ 1995); 
this second set was supplemented by additional {\em IUE} data 
that we will disregard in this study as they are far less suitable 
than the {\em HST} data for the present purposes. Both UV 
data sets were recovered from the International AGN Watch 
website. The {\em IUE} spectra used were those processed 
with the {\em IUE} New Spectral Image Processing System
(NEWSIPS; Nichols et al.\ 1993). The {\em HST} 
spectra were processed as described by Korista et al.\ 
(1995).

In addition to the UV spectra, there are 1248 optical 
spectra that were obtained beginning in 1988 (i.e., concurrently 
with the {\em IUE} spectra described above) for 13 years 
through 2001 (Peterson et al.\ 2002 and references therein). 
However, only a relatively small subset of these data are 
suitable for studying variability of the optical Fe\,{\sc ii} 
blends; the Fe\,{\sc ii} blends are broad, low-contrast features 
that can be measured or modeled usefully only in high-quality 
spectra. Also, as discussed below, we find that Fe\,{\sc ii} 
variability is most reliably measured by combining spectra to 
form difference spectra (i.e., subtraction of a low flux state 
spectrum from a high flux state spectrum) or rms spectra in 
order to isolate the parts of the spectrum that are varying. Since 
the amplitude of variability is small and since combining 
spectra propagates noise, it is essential to start out with 
high-quality data. Specifically, we require high signal-to-noise 
ratios, reasonably high spectral resolution in order to detect 
structure in the Fe\,{\sc ii} blends, and high-fidelity relative 
flux calibration over an extended wavelength range in order
to constrain the underlying continuum level. Moreover, 
the necessity of combining spectra requires that we select a 
homogeneous subset of the data. Given these requirements, the 
best subset is drawn from those obtained with the Lick 
Observatory 3-m telescope by A.V.\ Filippenko and 
collaborators (``Set H'' described by Peterson et al.\ 2002 and 
references therein). We selected from this subset all high-quality 
spectra that cover the observed wavelength range 
4000--6000\,\AA. These criteria yielded 73 spectra spanning 
the entire 13 years of the International AGN Watch program in 
a fairly representative way. These 73 spectra were put on a 
common flux scale by assuming that the narrow-line 
[O\,{\sc iii}]\,$\lambda\lambda4959$, 5007 fluxes are constant with 
$F(\mbox{\rm [O\,{\scriptsize III}]})\,\lambda5007 = 5.58 
\times 10^{-13}$\,ergs\,sec$^{-1}$cm$^{-2}$
(Peterson et al.\ 1991). This was accomplished using a version 
of the van Groningen \& Wanders (1992) flux scaling 
algorithm. After scaling each spectrum in flux, we verified the 
consistency of the calibration by measuring the continuum flux 
at 5100\,\AA\ and the total \Hbeta\ flux in each spectrum. 
Figure 1 shows a comparison of the continuum and \Hbeta\ 
fluxes measured here with published fluxes based on all of the 
International AGN Watch data. Agreement appears to be 
excellent. Figure 1 also underscores the point that the spectra 
we have selected for study are a suitably representative subset of 
the entire database.

\subsection{Fe\,{\small\bf II} Templates}

As noted above, the complex atomic structure of Fe$^+$
produces a large number of transitions that pervade the UV, 
optical, and IR spectrum. Doppler broadening within the BLR 
causes the Fe\,{\sc ii} emission lines to blend together and with 
lines of other species, thus making it extremely difficult to 
accurately measure the Fe\,{\sc ii} emission-line flux. 
However, it was first noted by Phillips (1977, 1978) that the 
optical Fe\,{\sc ii} spectra of  Seyfert 1 galaxies are very 
similar, differing from one AGN to the next primarily in the 
amount of Doppler broadening. In this context, the Seyfert 1 
galaxy I Zw 1 (Sargent 1968; Phillips 1976) is of particular 
interest: the widths of the Fe\,{\sc ii} and Balmer lines in this 
source are only $\sim1100$\,\kms, among the very narrowest 
permitted lines observed in Seyfert 1 spectra. Phillips 
demonstrated that by suitably broadening the I Zw 1 spectrum 
to match the 
broad \Hbeta\ widths of other Seyfert 1s, the observed 
Fe\,{\sc ii} profiles could be matched reasonably well. This 
suggested that the Fe\,{\sc ii} emission arises co-spatially with 
the broad \Hbeta\ emission. This also suggested to Boroson, 
Persson, \& Oke (1985) and Boroson \& Green (1992) that the 
broadened I Zw 1 spectrum could be used as a template to 
model the Fe\,{\sc ii} contribution to AGN emission-line 
spectra, although of course their interest was more focused on 
removing Fe\,{\sc ii} as an emission-line contaminant than in 
using it to determine the Fe\,{\sc ii} emission flux. Boroson \& 
Green used high-quality spectra of 
I Zw 1 to develop an Fe\,{\sc ii} emission-line 
template that could be used generally to remove Fe\,{\sc ii} 
emission from AGN spectra over the range 4250--7000\,\AA. 
Recently, a revised and improved optical Fe\,{\sc ii} template 
based on I Zw 1 and covering the range 3535--7534\,\AA\ has 
been prepared by V\'{e}ron-Cetty, Joly, \& V\'{e}ron (2004). It is this 
template, shown in Fig. 2,
that we use in analysis of the optical spectra of NGC 5548.
Templates with a range of intrinsic line widths were generated
by folding the entries of Fe\,{\sc ii} transitions listed in 
Table~A.1 of V\'{e}ron-Cetty et al.\ (their broad system L1) with 
Lorentzian\footnote{We follow here the authors on the choice of 
profile shape for the template. However, the actual profile shape 
has little importance in this context, especially once the template 
is broadened.} profiles of different FWHM. 

A similar template for the UV spectrum (1250--3090\,\AA), 
again based on I Zw 1, has been constructed from {\em HST} 
spectra by Vestergaard \& Wilkes (2001). We use this template,
shown in Fig.\ 3, 
to attempt to model the UV Fe\,{\sc ii} emission in NGC 5548. 
Vestergaard \& Wilkes present several versions of UV iron templates,
including separate templates of Fe\,{\sc ii} and Fe\,{\sc iii} emission, 
respectively, each of which have options of including or excluding
certain multiplets that are characteristically strong in I\,Zw\,1.
Two templates of UV iron emission were fit to the NGC 5548 data:
(1) a template consisting only of Fe\,{\sc ii} emission transitions,
excluding Fe\,{\sc ii} UV\,191 at 1786\,\AA, and
(2) a template consisting of both Fe\,{\sc ii} and Fe\,{\sc iii}
emission transitions, excluding Fe\,{\sc ii} UV\,191 and Fe\,{\sc iii} 
UV\,34 as these multiplets are particularly strong in I\,Zw\,1 and
the spectra of NGC\,5548 show little indication of their presence.
As will be clear from \S\,2.4, the template that includes
Fe\,{\sc iii} emission provides a poor fit to the data.
We note that the Fe\,{\sc ii} emission shortward of the C\,{\sc iv} 
emission line tends to be overpredicted by using 
the template as originally presented by Vestergaard \& Wilkes. 
Forster et al.\ (2001) came to a similar conclusion for a small 
fraction of the spectra in their sample.
The bluest part of the {\it HST} spectrum of I\,Zw\,1 (below $\sim$1500\,\AA{})
on which the UV template is based
was not observed at the same time as the longer-wavelength segments
and I\,Zw\,1 increased in luminosity between these two observations.
Vestergaard \& Wilkes rescaled the shortest-wavelength segment to match the 
longer-wavelength part of the spectrum to produce a full-range UV template.
However, it was recognized then that the Fe\,{\sc ii} emission in the
template below $\sim$1550\,\AA\ may need to be rescaled for some 
objects.  By fitting the NGC\,5548 spectra we found that 
very reasonable fits can be obtained by applying a scaling factor 
of 0.25 to the template Fe\,{\sc ii} emission shortward of 
1550\,\AA.

The general fitting procedure used here is described by Vestergaard \&
Wilkes (2001; section 4.2 in that paper). 
A power-law continuum is subtracted from the spectrum, which makes 
the Fe\,{\sc ii} emission more visible, before fitting the 
Fe\,{\sc ii} templates.  The power-law continuum is
fitted to windows that are virtually free of line emission, namely
$\sim$4200\,\AA\ and $\sim$5650\,\AA\ in the optical and $\sim$1350\,\AA\
and $\sim$2000\,\AA\ in the UV.
An iteration over continuum setting and Fe\,{\sc ii} template fitting
may be necessary to get the best fit.  We consider the best template fit to
be that which leave the residual spectra virtually featureless in regions
dominated by Fe\,{\sc ii}.
Additional details on our fitting of the optical 
spectrum of NGC 5548 are given in \S{2.3.3.}

\subsection{Optical Fe\,{\small\bf II} Blends}

We made an initial attempt to fit each of the 73 selected spectra 
with the optical Fe\,{\sc ii} template. We found this to be very 
difficult because of blending of Fe\,{\sc ii} with other features 
in the spectra, notably weak narrow lines and small-scale 
structure due to features in the underlying galaxy spectrum. 
Given the problems we encountered, we had little confidence 
in these results.

\subsubsection{Difference Spectra}

We therefore began experimenting with alternative approaches 
that would mitigate some of the problems due to contaminating 
features. One method is to produce difference spectra, i.e., 
to subtract a low flux-state spectrum from a high flux-state 
spectrum. This effectively removes all the non-variable 
components from the spectrum, including narrow emission 
lines and host galaxy light, although sometimes small residuals 
of these features can be seen, usually as a consequence of 
seeing and aperture effects (e.g., Peterson et al.\ 1995) or small 
differences in the line-spread function between pairs of 
observations. A potential liability of this approach is that the 
uncertainties in difference spectra can be fairly large; the 
differences between two spectra are often quite small, but the 
uncertainties add in quadrature. We attempted to mitigate this 
problem by making a high signal-to-noise ratio low-flux state 
spectrum by co-adding a number of individual faint-state 
spectra, and then subtracting this ``standard low-state spectrum'' 
from all of the individual spectra. To form a standard low-state 
spectrum, we identified several spectra obtained during the 
faintest periods in the 13-year monitoring program. The 
standard low-state spectrum as formed by averaging 3 spectra 
from the fourth year of the AGN Watch program (specifically, 
the spectra obtained on Julian dates 2448733, 2448765, and 
2448810) with two spectra obtained during the thirteenth year 
(on Julian dates 2452030 and 2452045).

Use of difference spectra greatly improved our ability to model 
the optical Fe\,{\sc ii} emission with a template, although the 
results were still not quite satisfactory. We found that there was 
degeneracy at about the 10\% level in fitting the Fe\,{\sc ii} 
fluxes (i.e., changes in flux of 10\% made no discernible 
difference in the overall quality of the fit). Moreover, 
comparison of the inferred Fe\,{\sc ii} fluxes between 
observations that are closely spaced in time often showed large 
differences; indeed, if the errors are random, we would infer 
that the level of precision of the Fe\,{\sc ii} fluxes is only 
$\sim30$\%, about an order of magnitude more uncertain than 
our \Hbeta\ fluxes (cf.\ Peterson et al.\ 2002). 
Despite these large uncertainties in individual spectra, we 
nevertheless find a clear correlation between the continuum 
flux at 5100\,\AA\ and the Fe\,{\sc ii} flux. 
We find a Pearson's linear correlation coefficient of 0.73 
and a probability of a chance (or null) correlation below 
0.0001.  While we caution about placing too much emphasis 
on the specific numbers, these results emphasize the reality
of such a correlation.

These measurements establish that the optical Fe\,{\sc ii} 
fluxes do indeed vary in a way that is correlated with the 
continuum. However, we are unable to establish a timescale for 
the response of the optical Fe\,{\sc ii} blends to continuum 
changes, i.e., to measure a reverberation lag. First, the 
sampling of these 73 spectra is comparatively sparse, as can be seen 
clearly in Fig.\ 1. The median interval between observations is 
nearly 60 days. For reference, cross-correlation of the 73 optical continuum 
and \Hbeta\ fluxes measured from these spectra and displayed 
in Fig.\ 1 yield time delays $\tau_{\rm cent} = 5.3 \pm 11.8$ 
days and $\tau_{\rm peak} = 17 \pm 4$ days, where these two 
quantities are respectively the centroid and peak of the optical 
continuum/\Hbeta\ cross-correlation function; both of these 
measurements are consistent with the result obtained using all 
the AGN Watch data shown in Fig.\ 1,
although the centroid value is formally highly uncertain. Second, 
the very large uncertainties in the optical Fe\,{\sc ii} fluxes obviate any 
chance of obtaining even a marginal time-lag measurement. Formally,
cross-correlation of the continuum and Fe\,{\sc ii} light curves
yields $\tau_{\rm cent} = 12 \pm 27$ days and 
$\tau_{\rm peak} = 11 \pm 34$ days
with a peak correlation coefficient $r_{\rm max} = 0.75$.
We can conclude, however,
that the lag for the optical Fe\,{\sc ii} emission is less than
several weeks; this is not an especially stringent upper limit
as it exceeds the value of any other reliably measured lag in
NGC\,5548.

Our original planned strategy had been to start 
with the most suitable optical spectra and progress to the less 
suitable, but better time-sampled data. However, we abandoned 
that approach when these initial results made it clear that 
measuring the Fe\,{\sc ii} fluxes accurately even in the best 
data is extremely difficult; measuring the Fe\,{\sc ii} fluxes in 
less suitable data would not improve the situation.

Inspection of the fits to our 73 difference spectra indicate that the 
difficulties in accurately fitting Fe\,{\sc ii} templates to the 
spectra are largely attributable to noise in the spectra, of both 
Poisson and non-Poisson origin. In other words, we might hope 
to obtain more accurate flux measurements by combining not 
only low-state, but high-state spectra as well. This proved to be 
essential for estimating the amplitude of Fe\,{\sc ii} flux 
variability. In this case, we examined the continuum flux 
measurements for all the AGN Watch data 
in Fig.\ 1 and computed their mean and standard 
deviation. We defined a mean high-state spectrum by 
combining all of the spectra in our selected sample of 73 spectra
for which the continuum flux is 
more than one standard deviation above the mean value. These 
spectra are those obtained on Julian dates 2448243, 2448280, 
2449802, 2449830, 2449930, 2449986, 2450336, 2450361, 
2451017, 2451056, 2451077, and 2451221. Similarly, we 
defined a mean low state spectrum by combining all of the 
spectra for which the continuum flux is more than one 
standard deviation below the mean values; the spectra that meet 
this criterion were obtained on Julian days 2447912, 2448089, 
2448102, 2448132, 2448733, 2448746, 2448765, 2448798, 
2448837, 2450664, 2452030, and 2452045. We then computed 
the difference between these two in order to produce a high 
signal-to-noise ratio difference spectrum, as shown in Figs.\ 4
and 5.

\subsubsection{Mean and RMS Spectra}

Another approach to more clearly isolating the variable part of 
the spectrum (and thus suppressing the constant components 
that contaminate the optical Fe\,{\sc ii} spectrum) is to 
compute a root-mean-square (rms) spectrum.
Using all of the 73 spectra in the sample,
we formed a mean spectrum,
\begin{equation}
\label{eq:meanspec}
\overline{F(\lambda)} = \frac{1}{N} \sum_{i=1}^N 
F_i(\lambda),
\end{equation}
where $F_i(\lambda)$ is the $i$th spectrum of the
$N$ spectra that comprise the database.
Similarly, the rms spectrum is defined by
\begin{equation}
\label{eq:rmsspec}
S(\lambda) = \left[ \frac{1}{N - 1} 
\sum_{i=1}^N \left( F_i(\lambda) - \overline{F(\lambda)} 
\right)^2 
\right]^{1/2}.
\end{equation}
The mean and rms spectra formed from our 73 optical spectra 
are shown in Fig.\ 6.
This mean spectrum is the only spectrum which we fitted for this
discussion that does not strictly contain variable emission
components only. The mean spectrum was fitted to allow assessment
of the relative Fe\,{\sc ii} variability amplitude.

\subsubsection{Identifying the Best Fitting Method of the Optical Iron 
Emission in NGC 5548}

The 4450--4650\,\AA\ and 5050--5600\,\AA\
regions of AGN and quasars are commonly dominated by \feii{} emission
which can be reasonably well-fitted
by an optical \feii{} template generated from the I Zw 1 emission
spectrum (Boroson \& Green 1992; V\'{e}ron-Cetty et al.\ 2004).
However, we were unable to fit these regions in the NGC\,5548 optical 
spectra with either the V\'{e}ron-Cetty et al.\ or the Boroson \& Green 
templates in their original form, i.e.,
with relative line intensities as observed in I\,Zw\,1. The top
panel of Fig.\ 7 illustrates the result of 
accounting for all the flux at $\sim4500$\,\AA\
above a power-law continuum with the V\'{e}ron-Cetty et al.\
\feii\ template. Clearly, this strongly overpredicts the \feii\ 
flux in the 5050--5700\,\AA\ range. Fitting the template to match
the 5050 -- 5600\,\AA{} \feii\ features instead leaves strong 
residuals at $\sim4500$\,\AA.

We examined three possible solutions to this problem. First, the \feii\
features may not have the relative line intensity as observed for
I Zw 1. Second, the emission at $\sim$4500\,\AA\ may be due to
extended wings of some combination of \Hgamma, 
He\,{\sc ii}\,$\lambda4686$, and \Hbeta.
Third, additional line emission, previously unaccounted for, 
may be present. We address each in turn in the following.

\begin{enumerate}
\item 
We did indeed find that the spectra could be fitted well if the 
5050--5600\,\AA\ \feii\ features in the template are rescaled. The
best fits to the mean, difference, and rms spectra of such 
modified V\'{e}ron-Cetty et al.\ (or Boroson \& Green) templates 
yields scalings relative
to (fixed) $\sim$4500\,\AA{} \feii\ strength of $\sim$0.4, 0.25, 0.2,
respectively. However, such low intensity ratios are not commonly
seen in AGN spectra; relative intensities of 0.7 -- 0.8 seem more
typical. Also, it is apparently difficult to obtain very low
intensity ratios of these lines in standard photoionization equilibrium
modeling (M.\ Joly 2004, private communication). 
This solution is hence not particularly attractive.

\item To test if extended wings of the \Hgamma, \heii, and \Hbeta\
lines can account for the $\sim4500$\,\AA{} residual emission
seen in the lower panel of Fig.~7 we fitted Gaussian functions
to \Hgamma{} and \heii\,$\lambda4686$, while a combination of Gaussian and 
Lorentzian functions was necessary to reproduce the (variable) 
\Hbeta{} profile.  These modeled lines were then subtracted 
from the data.  The results are shown in Fig.\ 8.
The upper panel shows the best-fit width to the He\,{\sc ii} line
of 10,000\,\kms (FWHM). A strong residual feature at 4500\,\AA\ shows
that we still cannot account for all the flux in this region.
A second attempt with a broader (12,000\,\kms\ FWHM) \heii\ line,
as shown in the lower panel of Fig.\ 8, results in very little
improvement.

\item
An alternative explanation is that additional 
previously unidentified line emission is
present in the NGC 5548 spectrum.
Although emission lines other than \feii\ are rarely 
seen at $\sim4500$\,\AA{} in quasars (e.g., Boroson \& Green 1992),
NGC 5548 may have a strong contribution of \hei\
$\lambda$4471, given the presence of strong \hei\ $\lambda$5876
observed at the edge of the spectra (Figs.\ 4, 5, and 6). In fact, 
the presence of \hei\,$\lambda$5876 also implies the existence 
of the \hei\ singlet lines at 4922\,\AA{} and 5016\,\AA. 
Common intensity ratios to \hei\ $\lambda$\,5876 are 0.3, 0.1, 
and 0.1 for the 4471\,\AA, 4922\,\AA, and 5016\,\AA{} lines, respectively, 
although the 4471\,\AA{} intensity ratio may be as low as 0.1 
(K. Korista 2004, private communication). 
Both panels of Fig.\ 8 illustrate that the excess emission at 
$\sim4500$\,\AA{} remaining when subtracting other known 
sources of line emission is consistent with the presence of 
\hei\ $\lambda$4471 (thin solid curve) at an intensity of 
0.3 times that of \hei\ $\lambda$5876. 
\end{enumerate}

These arguments collectively justify our inclusion in our general 
fitting of the non-apparent \hei\ $\lambda \lambda$4471, 4922, 5016 
lines with the above-mentioned intensity ratios as found in 
photoionization equilibrium modeling.
Figure 5 illustrates the method we thereafter adopted for fitting
the \feii\ emission: the expected \hei\ features, based on the
observed 5876\,\AA\ line, were subtracted before fitting the
\feii\ emission. The adopted template fitting of the difference,
mean, and rms spectra are shown in Figs.\ 4 and 6.

We note that both the broad \hei\,$\lambda$5876 and \heii\,$\lambda$4686
features are observed shifted to the blue by $\sim$1000\,\kms\ to
1500\,\kms{}; this is apparent even in the observed spectra.
It is particularly clear in the mean spectrum (Fig.\ 6) where the
narrow \hei\,$\lambda$5876 component is observed at 5876\,\AA{}
yet the peak of the broad component is blueshifted; we reproduced
this shift when modeling and subtracting the expected \hei\ lines 
in the mean spectrum. 
Interestingly, there is no clear indication that the (broad)
\heii\,$\lambda$1640 line is shifted by 
such a large amount.

Upon subtraction of the best-fit \feii\ templates, a few small residuals 
may still remain, especially in the mean spectrum (Fig.\ 6) where 
some non-variable possibly narrower \feii\ transitions may be 
present; we did not attempt to include any transitions of the 
narrower \feii\ system identified by V\'{e}ron-Cetty et al.\ (2004), 
on the assumption that these would not appear in our difference
or rms spectra, which appears justified by Figs.\ 4 -- 6.
It is noteworthy that many of the smaller-scale features seen 
atop the 4450--4750\,\AA\ emission blend in the mean spectrum
are consistent with the expected positions of \feii\ m38.

\subsubsection{Results}

Our analysis of the optical Fe\,{\sc ii} data on NGC 5548 leads 
us to the conclusion that the Fe\,{\sc ii} emission is indeed 
variable in flux and that these variations are positively 
correlated with continuum variations. We are unable, however, 
to make any meaningful determination of any time delay 
between the continuum and Fe\,{\sc ii} emission blend 
variations.

In an attempt to assess the amplitude of 
variability of
the optical Fe\,{\sc ii} emission, we integrated the Fe\,{\sc ii}
flux in the best-fit template over the range 4250--5710\,\AA\
and divided it by the \Hbeta\ broad-line flux as measured in
the mean, difference, and rms spectra, respectively. The results are shown
in Table 1. Comparison of the Fe\,{\sc ii}/\Hbeta\ flux ratio 
in the variable spectra with the mean spectra shows that this
ratio is only 25--50\% smaller in
the variable spectra. This implies that the
optical Fe\,{\sc ii} emission varies with an amplitude 
of 50--75\% 
that of the broad \Hbeta\ component. The reason 
that \feii\ emission is not always obvious in rms spectra
is simply because the variations are 
generally rather small, and the low-contrast Fe\,{\sc ii} features 
are weak relative to \Hbeta\ and do not show up very well even in 
high-quality difference or rms spectra.

The best fit we achieve is with an optical Fe\,{\sc ii} template
broadened to a width similar to that of the variable \Hbeta{} line;
the \Hbeta{} line width in the rms emission-line spectrum shown in
the lower panel of Fig.\ 6 is about 6250\,\kms. Interestingly, this
template gives only a slightly improved fit to the 5050 -- 5600\,\AA{}
Fe\,{\sc ii} features relative to templates of somewhat broader widths 
of $\sim$10,000\,\kms; the differences in the fit and residuals are
practically insignificant.  

It is also worth 
noting, however, that for broadening larger than 
$\sim9000$\,\kms, the templates become essentially 
degenerate, i.e., it is hard to distinguish among them.

Given that reasonable fits are obtainable 
with a range of template line widths,
we carried out additional tests to try to determine the
maximum and minimum line widths for which the template
fitting process gives unacceptable 
results. We find a range of template line widths that yield 
``acceptable'' fits, and a smaller range that are ``preferred,'' 
for which even better fits are obtained. The degeneracy is 
mainly due to the flux density uncertainties in the difference 
spectra (section 2.3.1) 
and the relatively minute differences in the detailed shapes of
Fe\,{\sc ii} templates that differ by 250--750\,\kms{} in line 
widths, especially for large line widths (above about 6000\,\kms{}).
For the optical difference and rms spectra, which are not contaminated
by non-variable emission, we find that the template fits are 
unacceptable at line widths less than 4500\,\kms{}; ``preferred'' fits 
were obtained for widths above 5750\,\kms{}, and 
our ``best fit'' value of 6250\,\kms{} is a subjective 
assessment that this provides the best overall fit to the spectrum.
Establishing an upper limit on the broadening is more difficult
because of the stronger degeneracy of the templates 
with line widths above about 9000\,\kms. We are unable to place 
any meaningful constraints on the upper limit to the Fe\,{\sc ii} 
line widths.

Subtraction of the optical Fe\,{\sc ii} template leaves 
He\,{\sc ii}\,$\lambda4686$ 
in the residual spectrum. The width of this line in the 
residual spectrum is very broad, $\sim12,000$\,\kms, and the 
width is insensitive to which Fe\,{\sc ii} template is used, i.e., 
the difference between using the 6250\,\kms\ and 10,000\,\kms\ 
templates changes the He\,{\sc ii}\,$\lambda4686$  width 
by less than about 200\,\kms.
This He\,{\sc ii}\,$\lambda4686$ line width is 
very similar to the width of He\,{\sc ii}\,$\lambda1640$ 
measured in the Fe\,{\sc ii} subtracted UV difference
and rms spectra, discussed below, and to the 
He\,{\sc ii}\,$\lambda1640$ width
obtained during both UV monitoring campaigns, and 
significantly larger than the width of He\,{\sc ii}\,$\lambda4686$ measured in 
the rms spectrum, with no attempt to account for the 
possible effects of \feii, during the first year of the NGC 5548 monitoring 
campaign  (Peterson et al.\ 2004). 
It is not clear, however, 
whether the He\,{\sc ii}\,$\lambda4686$ 
width varies with time, as does, for example, 
the width of \Hbeta, or if the base of the He\,{\sc ii}\,$\lambda4686$ 
line gets broader when the Fe\,{\sc ii} is removed. Certainly, the 
responses of He\,{\sc ii}\,$\lambda1640$ and 
He\,{\sc ii}\,$\lambda4686$ are expected to be 
identical (Bottorff et al.\ 2002), and the lags are, in fact, 
consistent within their uncertainties (Peterson et al.\ 2004). 
Indeed, there seems to be no strong evidence that either the
lags or line widths of He\,{\sc ii}\,$\lambda1640$ and 
He\,{\sc ii}\,$\lambda4686$ are significantly 
different from
one another. The concerns of Bottorff et al.\ about the differences in 
the response of these two lines may be not as significant as 
previously thought.

\subsection{UV Fe\,{\small\bf II} Blends}

\subsubsection{UV Difference Spectra}

In order to determine whether or not the UV Fe\,{\sc ii} 
emission blends vary with the continuum, we carried out an 
analysis similar to that described above for the 
optical emission lines, 
though in this case we describe only our investigation of 
difference spectra. For both UV data sets described in \S{2.1}, 
we computed mean high-state and mean low-state spectra, 
again using the criteria that these spectra have continuum fluxes 
that are one standard deviation above and below the mean flux, 
respectively. The {\em HST} high state spectrum included 
those obtained on Julian dates 2449121, 2449122, 2449123, 
and 2449124, and those that comprise the low-state spectrum 
are those obtained on 2449097, 2449098, and 2449100. These 
spectra and their difference are shown in Figs.\ 9 and 10. For the {\em 
IUE} data, the high-state spectrum was constructed from the 
spectra obtained on Julian dates 2447538, 2447543, 2447546, 
2447621, 2447625, 2447629, 2447633, 2447634, 2447637, 
2447638, 2447641, 2447645, and 2447649, and the low-state 
spectrum is based on the spectra obtained on Julian dates 
2447577, 2447581, 2447688, 2447692, 2447729, 2447733, 
2447737, 2447741, and 2447745. These spectra and their 
difference are shown in Figs.\ 11 and 12.

\subsubsection{Results}

We fit the difference spectra with two types of template, (1) a 
pure Fe\,{\sc ii} template, and (2) a template that includes both 
Fe\,{\sc ii} and Fe\,{\sc iii}. Both templates are from 
Vestergaard \& Wilkes (2001) and described in \S~2.2. 
While Fe\,{\sc ii} is clearly 
present in the difference spectrum, it is not clear that Fe\,{\sc 
iii} is present. If Fe\,{\sc iii} is present, it is very weak. We 
find that inclusion of Fe\,{\sc iii} is not necessary to model the 
spectrum, and the spectrum is better fit by the pure Fe\,{\sc ii} 
template. 
One possible exception is the bump at $\sim$2200\,\AA, which is 
most likely Fe\,{\sc iii} (Vestergaard \& Wilkes 2001). However, 
the results of this study are not sensitive to the strength of
this feature  and we will not consider it any further.
Compared to the {\em HST} spectrum,
the {\em IUE} spectrum is harder 
to fit because it is noisier and covers a shorter wavelength 
range. On the other hand, if we constrain the slope 
of the {\em IUE} difference spectrum by using the 
slope measured in the {\em HST} difference spectrum, we obtain a 
satisfactory fit. We find that a better fit is obtained with 
the 10,000\,\kms\ template than for the 6250\,\kms\ template
(Figs.\ 10 and 12).
In particular, we find unacceptable fits using templates with
broadenings less than 8250\,\kms. Given the same issues pertaining
to the optical spectra, we are unable to place any upper limits
on template broadenings which produce acceptable fits.

We also compare the strengths of the UV Fe\,{\sc ii} blends
(integrated across the two intervals 1430--1500\,\AA\ and
1685--1850\,\AA)
and \Hbeta\ in the variable part of the spectrum with what is
observed in the mean spectrum, as we did with the optical Fe\,{\sc ii}
emission above\footnote{The \Hbeta\ fluxes were measured in mean or
difference spectra constructed from optical spectra that are
contemporaneous with the UV spectra used.}. 
The results of this comparison are given in 
Table 2. We see here that
the UV Fe\,{\sc ii}/\Hbeta\ flux ratio is significantly larger
in the variable part of the spectrum than in the mean spectrum
(we note in passing that analysis of the UV mean and rms
spectra leads to the same conclusion).
Unlike the optical Fe\,{\sc ii}
emission, the UV Fe\,{\sc ii} emission seems to vary with
a larger amplitude than \Hbeta.

\section{DISCUSSION}

The principal result of this study is that the optical and UV 
blends of Fe\,{\sc ii} emission in the spectrum of NGC 5548 
do indeed vary with time. These variations are correlated with the 
continuum variations. The optical Fe\,{\sc ii} blends vary with
an amplitude 50--75\% that of \Hbeta, and the UV 
Fe\,{\sc ii} blends seem to vary with an even larger amplitude
than \Hbeta.
The variations of these low-contrast features are
most clearly seen in difference or rms spectra, which remove 
non-variable structure in the spectra. However, the variable 
Fe\,{\sc ii} features are not always obvious in difference or rms 
spectra simply because these features are 
often lost in the noise. 
Previous claims of the lack of variability of Fe\,{\sc ii} seem to 
be consistent with this conclusion, given the sensitivity of the 
observations. Unfortunately, on account of the weakness of these 
features in difference or rms spectra, we are unable to determine 
the time scale for response of the Fe\,{\sc ii} emission lines to 
continuum variations. 

We note in passing that very  broad Fe\,{\sc ii} features
produce a ``pseudo-continuum'' (e.g., Wills et al.\ 1985)
that is not readily identified as line emission.
For example, Figs.\ 9 and 11 (upper panels) show that the 
1700--1900\,\AA\ region in NGC 5548 is relatively flat, 
giving the appearance that no (variable) emission lines are present,
even though this region is at a somewhat higher level than the
continuum at $\sim$1450\,\AA\ and 2025\,\AA.
However, the presence of Fe\,{\sc ii} emission 
in this spectral range is clear when fitting 
the spectrum with an Fe\,{\sc ii} template.

Interestingly, we find that the UV Fe\,{\sc ii} features in the 
difference or rms spectra are best fitted with templates that 
have a large velocity width, $\sim$10,000\,\kms; the optical
Fe\,{\sc ii} features are consistent with both such a broad 
template and one of the variable \Hbeta\ line width.
We have no ready explanation for this result, although
we might speculate that it may be related to the issues
responsible for the difficulty of accounting for the strength of 
the Fe\,{\sc ii} (Baldwin et al.\ 2004), including the unusual 
smoothness of the \feii\ emission; either microturbulence or 
velocity shear seems necessary to account for the
discrepancy between theory and observation. 

The issue of whether or not the optical and UV Fe\,{\sc ii} is
emitted from the same region is important for theoretical modeling
of their relative emission strengths (Baldwin et al.\ 2004).
While the errors on the Fe\,{\sc ii} emission-line response
time are very large, it is possible to constrain the size of
the Fe\,{\sc ii}-emitting regions indirectly through the
line-broadening parameters under the assumption that
Fe\,{\sc ii} follows the same virial relationship
seen in every other well-characterized variable line
in NGC~5548 (Peterson \& Wandel 1999). If we assume
that the virial product $c\tau_{\rm cent} {\rm FWHM}^2/G$
is the same for all lines in NGC 5548, using measurements
from Peterson et al.\ (2004), we can conclude that
the optical Fe\,{\sc ii} emission, with
${\rm FWHM} > 4500$\,\kms, arises at
$r < 24$ light days, and the UV Fe\,{\sc ii}
emission, with ${\rm FWHM} > 8500$\,\kms,
arises at $r < 7$ light days.
This limit on the size of the optical Fe\,{\sc ii} emission 
gas does not violate 
the weak constraint on the time delay of several weeks based 
on cross-correlation of the continuum and optical Fe\,{\sc ii}
emission light curves (section 2.3.1). 
Given the assumptions, this is a weak conclusion, but
it is worth noting that differing line widths for the
UV and optical Fe\,{\sc ii} emission suggests that the UV and
optical Fe\,{\sc ii} might be emitted predominantly
from different regions within the BLR.
If indeed the BLR gas density increases steadily and smoothly
toward the center of the BLR, the distance limits are broadly 
consistent with expectations from photoionization models 
(e.g., Fig.\ 11 of Baldwin et al.\ 2004), namely that UV Fe\,{\sc ii} 
is most efficiently emitted from higher-density gas (presumably 
located closer to the central source), while optical Fe\,{\sc ii} 
is more efficiently emitted from lower-density gas farther out.

V\'{e}ron-Cetty, Joly, \& V\'{e}ron (2004) find
that the I\,Zw\,1 Fe\,{\sc ii} emission spectrum 
is comprised of two components of different widths,  
$\sim$1100\,\kms{} and $\sim$300\,\kms{}, the latter mostly being
forbidden iron transitions. 
This suggests that some of the Fe\,{\sc ii} emission originates in 
the narrow-line region. The 
fact that no variations are detected in the narrow Fe\,{\sc ii}
features by virtue of their absence in the residual spectra of
the difference and rms spectra is consistent with this explanation;
blends of narrow Fe\,{\sc ii} lines may be present in the mean
optical spectrum, especially longward of [O\,{\sc iii}].
We cannot identify narrow Fe\,{\sc ii} in any of the
UV spectra we have examined.

Our conclusion that the amplitude of optical Fe\,{\sc ii} variability 
is somewhat smaller than that of \Hbeta, but is often of too low 
contrast for reliable detection, seems consistent with results for 
other sources (Kollatschny et al.\ 1981; Kollatschny \& Fricke
1985; Giannuzzo \& Stirpe 1996; Doroshenko et al.\ 1999).  Apparent
exceptions are Fairall 9 and NGC 7603. Kollatschny et al.\ (2000)
found the optical Fe\,{\sc ii} to vary as strongly as \Hbeta{} for 
NGC 7603.  In the case of Fairall 9,
Kollatschny \& Fricke find that both \Hbeta\ and the optical
Fe\,{\sc ii} emission vary with large amplitudes, by factors
of 5.6 and 2.6, respectively. For more direct comparison with
this result, we restricted our Fe\,{\sc ii} measurement window
to the 4570\AA, 4928\,\AA, and 5018\,\AA\ multiplets, the
approximate region used by Kollatschny \& Fricke, and found
similar values to those we quote in Table 1. 
We conclude that the relative amplitudes of 
Fe\,{\sc ii} and \Hbeta\ flux variability amplitudes can 
differ significantly among AGNs.
 
As current photoionization models fall short in reproducing the 
observed AGN Fe\,{\sc ii} emission, Collin \& Joly (2000) 
suggest that Fe\,{\sc ii} in AGNs is due entirely to 
shock (i.e., collisional) excitation. They draw parallels with 
strong Fe\,{\sc ii} emitting stellar sources which have no X-ray or
ionizing emission, but do have strong variability and strong outflows.
Baldwin et al.\ (2004) do find that collisional excitation can explain the
observed Fe\,{\sc ii}, though they are reluctant to 
consider introduction of another component, which produces no
other significant emission, into an already complex
picture of the BLR. Moreover, it is not clear 
how such a model could reproduce the observed correlation between
the optical Fe\,{\sc ii} and the continuum fluxes that we observe.

The template fits show that the I\,Zw\,1 Fe\,{\sc ii} templates are 
generally reasonably good  matches to the optical and UV Fe\,{\sc ii} 
emission from NGC\,5548. The exceptions are, as noted earlier, the 
apparent absence of the Fe\,{\sc iii} transitions over the
range 1825--2100\,\AA\ and the poor match of the original 
templates at 2025\,\AA\ where the 
Fe\,{\sc ii} emission appears to be missing.

\section{CONCLUSIONS}

Our chief conclusions of our study of the Fe\,{\sc ii}
emission in NGC\,5548 can be summarized as follows:

\begin{enumerate}
\item
The optical and UV Fe\,{\sc ii} emission blends do indeed vary in response to 
variations in the continuum emission,
i.e., the Fe\,{\sc ii} and continuum flux variations are strongly correlated.
Detection of these variations is difficult because these lines
are very broad and of low contrast.

\item
The optical Fe\,{\sc ii} varies in strength with an amplitude
of 50\% -- 75\% of that of \Hbeta.
The UV Fe\,{\sc ii} varies much more strongly.

\item
These results are consistent with most earlier studies claiming no
variability owing to the difficulty of detecting small-scale
variations in low-contrast features.

\item
Unfortunately, the weakness of the Fe\,{\sc ii} features preclude 
a meaningful variability time scale of this emission to be obtained.
The flux variations of the optical Fe\,{\sc ii} blends lag behind
those in the continuum by no more than several weeks.  

\item
For reasons that are unclear, we find somewhat improved fits using 
UV Fe\,{\sc ii} templates with large widths of order 10,000\,\kms, 
as opposed to templates with widths similar to that of \Hbeta. 

\item
Fe\,{\sc ii} removal improves measurement of other emission lines.
The He\,{\sc ii}\,$\lambda$4686 feature in the iron-subtracted
spectrum displays a line width consistent with that of 
He\,{\sc ii}\,$\lambda$1640, as expected from photoionization models. 
Failure to account for the effect of Fe\,{\sc ii} 
emission seems to yield a 
He\,{\sc ii}\,$\lambda$4686 line width that is too small.

\item
Because we cannot meaningfully constrain either the line width or 
the reverberation response timescale for Fe\,{\sc ii}, we are unable 
to determine conclusively
whether or not the Fe\,{\sc ii} emission originates in
the same region as \Hbeta{} or whether the UV and optical \feii\ is
spatially co-emitted; this issue has importance for the
theoretical modeling of the relative UV and optical Fe\,{\sc ii}
emission strengths (e.g., Baldwin et al.\ 2004). 

However the line widths below which the optical and UV templates
produce unacceptable fits to the observed Fe\,{\sc ii} emission
do suggest that the UV and optical Fe\,{\sc ii} may originate in
different regions, although there may be some overlap of these 
regions given our inability to place meaningful upper limits on 
the Fe\,{\sc ii} line widths.

\end{enumerate}

\acknowledgements
The authors are grateful for support of this research by NASA 
through Grant No.
HST-AR-09549-01.A from the Space Telescope Science 
Institute, which is
operated by the Association of Universities for Research in 
Astronomy, Incorporated, under NASA Contract NAS5-26555,
and by NSF through grant AST-0205964.
Kirk T.\ Korista, Monique Joly, Mira V\'{e}ron-Cetty
and Philippe V\'{e}ron are thanked for helpful discussions
and comments on an earlier version of the manuscript.
We wish to thank Kirk Korista in particular for pointing
out the possible importance of the He\,{\sc i} lines.
Mike Goad is thanked for a helpful and constructive referee's
report.


%
\begin{figure}
\plotone{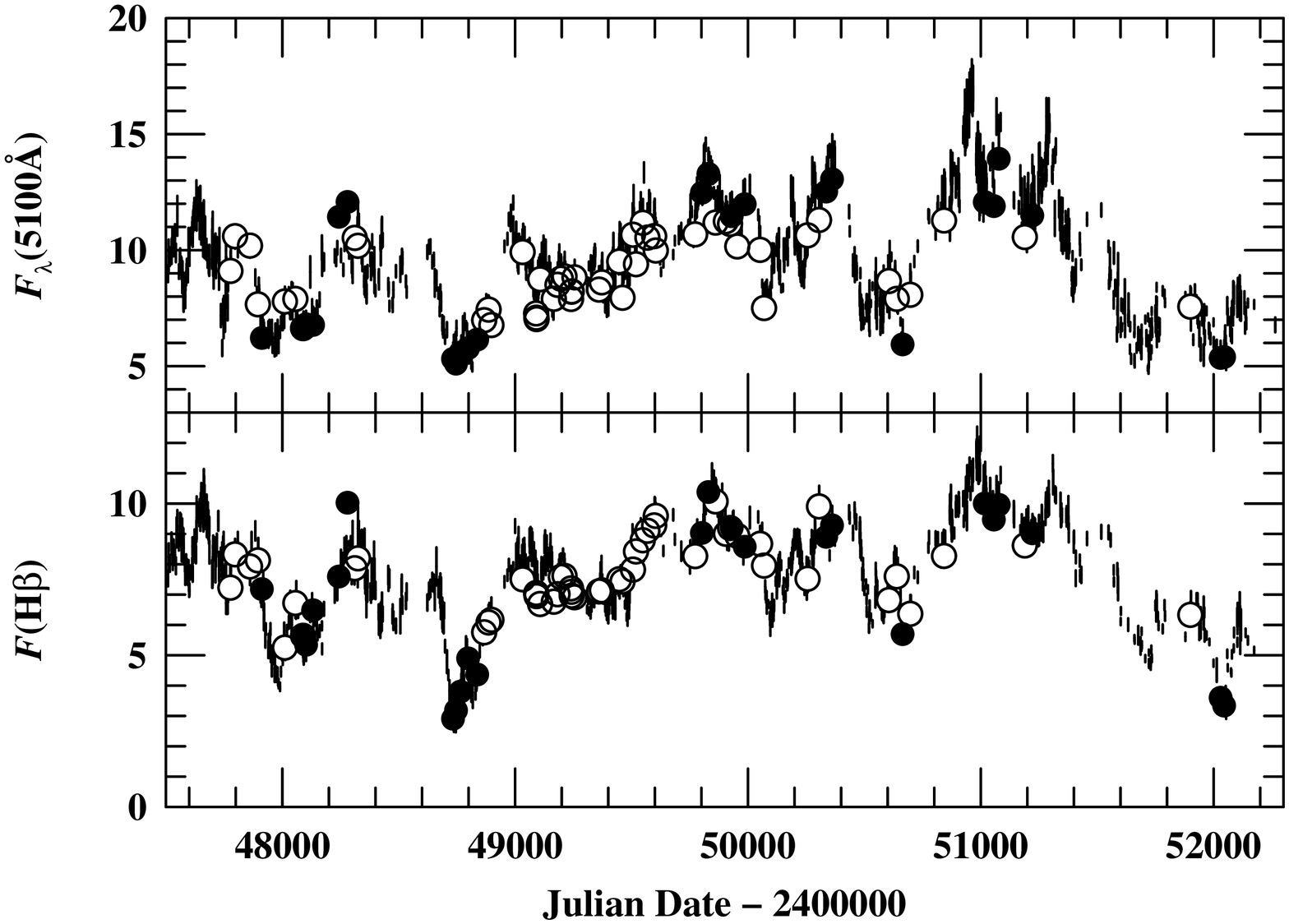}
\caption{Optical continuum (upper panel) and \Hbeta{}
(lower panel) light
curves for NGC 5548 for the period 1988 to 2001. The vertical
lines show the $\pm 1 \sigma$ error ranges for the light
curves based on the International AGN Watch database
(Peterson et al.\ 2002 and references therein), and the
open and filled circles show the measurements based on the 73 rescaled
spectra used in this study. 
The filled circles mark the epochs of
the data used to construct the low and high state spectra shown in
Figure 4 and described in section 2.3.1. 
Continuum fluxes are 
in units of $10^{-15}$\,\contunits, and \Hbeta\ fluxes
are in units of $10^{-13}$\,\lineunits. The distribution
of open and filled circles shows that this particular subset
of spectra represents the entire database very well.}
\end{figure}

%
\begin{figure}
\plotfiddle{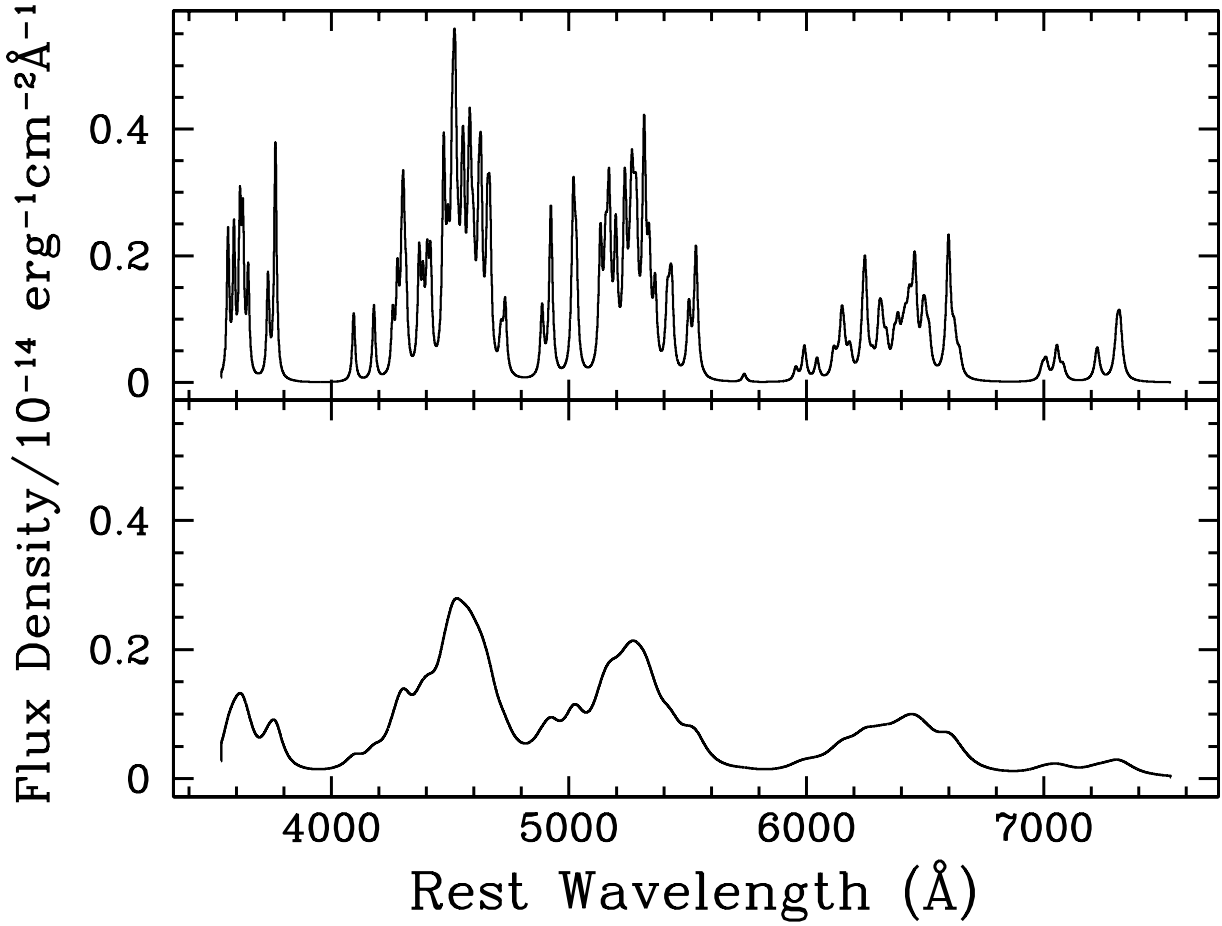}{50pt}{270}{358}{471}{0}{0}
\caption{The upper panel shows the optical Fe\,{\sc ii} template
constructed by V\'{e}ron-Cetty, Joly, \& V\'{e}ron (2004), 
 reproduced as described in \S\,2.2; the intrinsic line width is 
1000\,\kms\ (FWHM).
The lower panel shows this template 
with an intrinsic Doppler width of 6250\,\kms\ (FWHM).}
\end{figure}

%
\begin{figure}
\plotfiddle{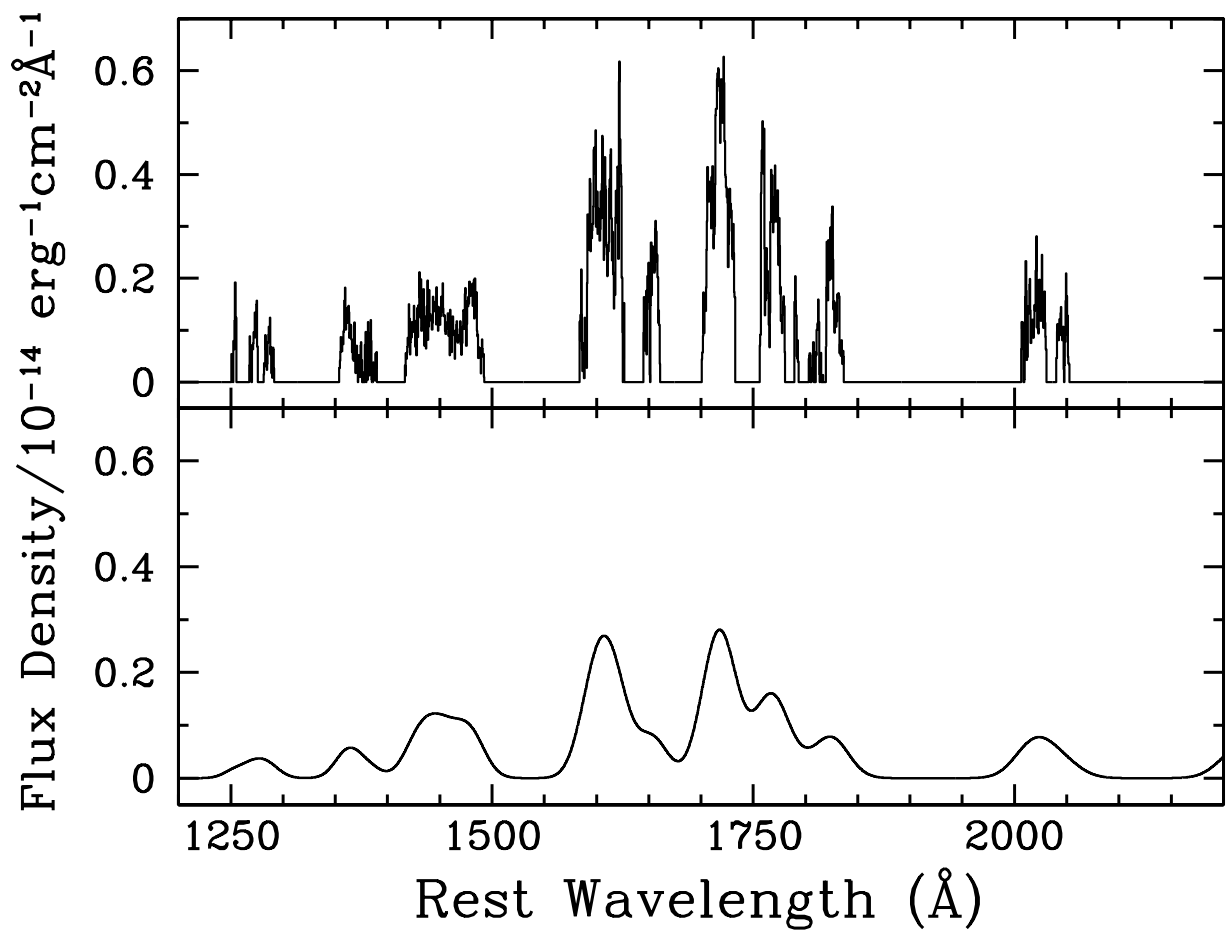}{50pt}{270}{358}{471}{0}{0}
\caption{The upper panel shows the unbroadened ultraviolet Fe\,{\sc ii} template
constructed by Vestergaard \& Wilkes (2001). 
The fluxes shortward of  1550\,\AA\ are scaled by a factor 0.25 compared
to the original template of Vestergaard \& Wilkes (see \S\,2.2),
which yields a better fit in general to the data analyzed here. 
Note that no Fe\,{\sc iii}
emission is present in this template.
The lower panel shows this template artificially broadened to
a Doppler width of 6250\,\kms\ (FWHM).}
\end{figure}

%
\begin{figure}
\plotfiddle{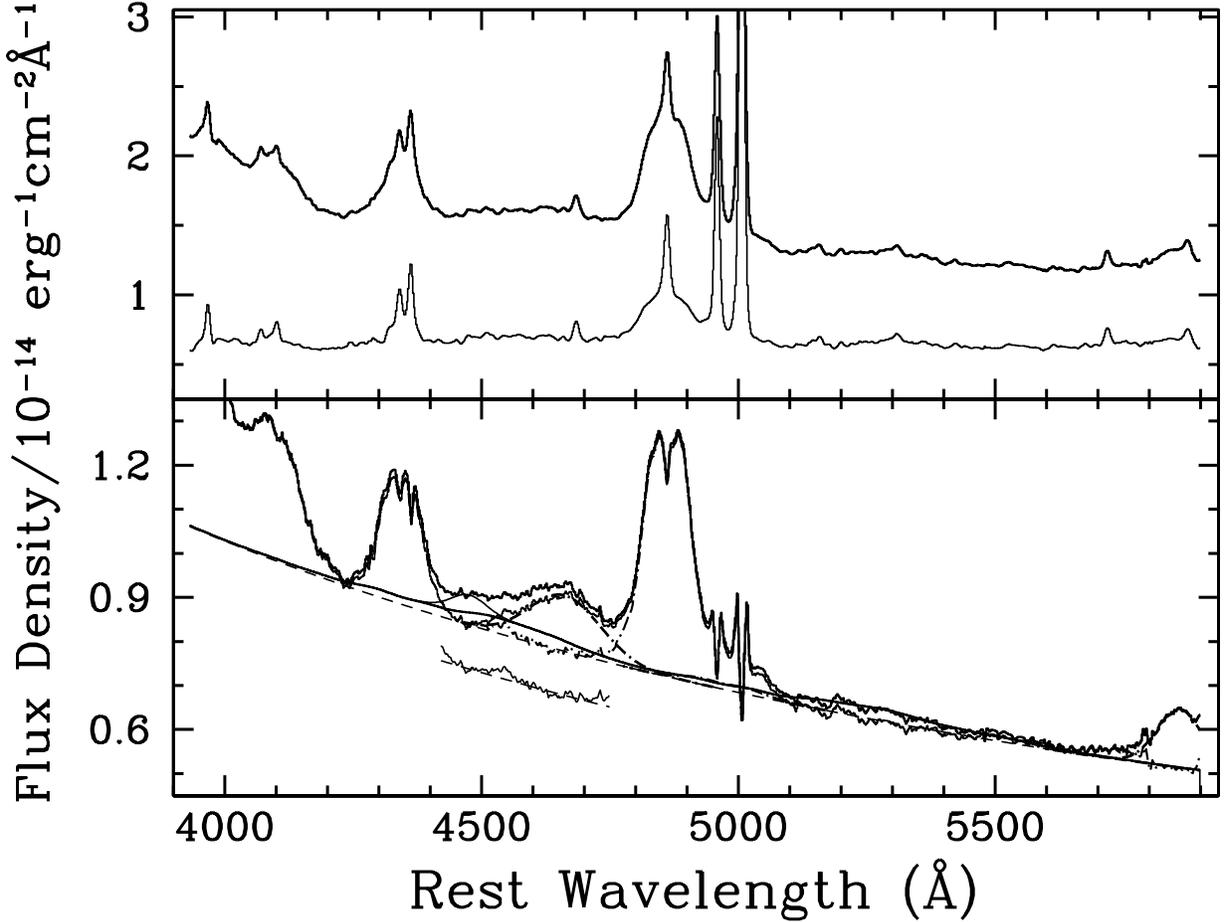}{50pt}{270}{358}{471}{0}{0}
\caption{The upper panel shows the high state and low state
optical spectra of NGC 5548 in the rest frame.
The lower panel shows the difference spectrum formed by
subtraction of the low state spectrum from the high
state spectrum (upper spectrum in the upper panel)
along with our preferred final fit to the spectrum,
as described in \S{2.3.3.}.
Note that the residuals in the 
[O\,{\sc iii}]\,$\lambda\lambda4959$, 5007 lines arise from
small differences in the line-spread function; the total 
line flux in the residuals is nevertheless zero.
The smooth solid thick line shows the broadened 
(FWHM = 6250\,\kms) and scaled Fe\,{\sc ii} template, 
the dashed line shows a fit to the underlying continuum,
and the dash-dot line shows a Gaussian fit 
($\sim$12,000\,\kms\ FWHM) to the residual 
He\,{\sc ii}\,$\lambda4686$ profile. 
The smooth solid {\em thin} line represents the 
He\,{\sc i}\,$\lambda\lambda$\,4471, 4922, 5016, 5876 lines;
for visibility, these lines are shown {\em on top} of the 
Fe\,{\sc ii} template fit.  The spectrum after subtraction 
of the Fe\,{\sc ii} template, the He\,{\sc i} lines, and 
the He\,{\sc ii} fit is also shown (dashed line). For visibility
a section of this residual spectrum and the underlying
continuum is shown shifted to a lower flux level.}
\end{figure}

%
\begin{figure}
\plotfiddle{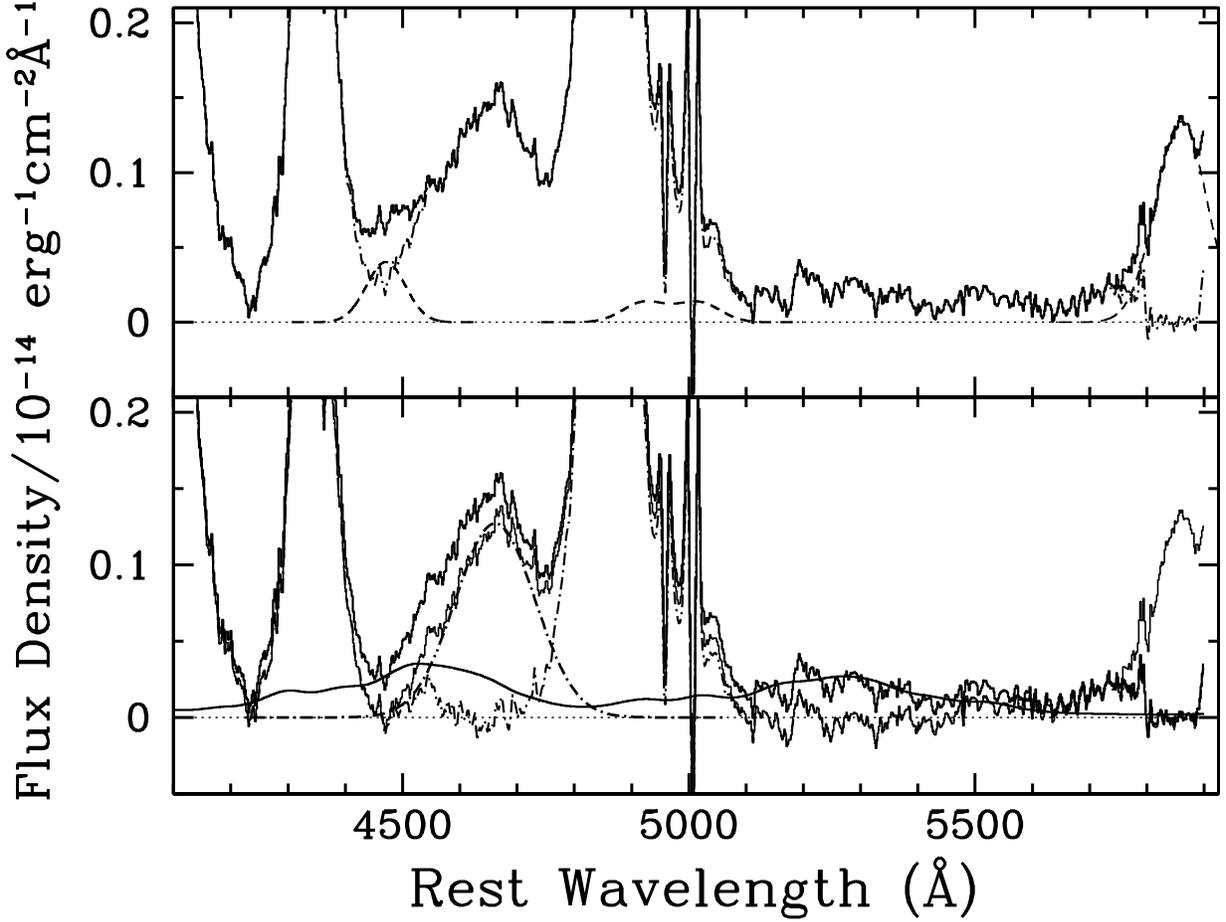}{50pt}{270}{358}{471}{0}{0}
\caption{
Illustration of our method for fitting the optical spectra of NGC 5548.
An enlargement of the difference spectrum from the
lower panel of Fig.\ 4 with the continuum subtracted 
is shown in the upper panel.
The He\,{\sc i}\,$\lambda \lambda$\,4471, 4922, 5016, 5876
lines (dashed) and the spectrum remaining after their 
subtraction (dot-dashed) are also shown.
The $\lambda \lambda$\,4471, 4922, 5016 profiles are those
expected based on the observed $\lambda$\,5876 line (see text).
In the lower panel the smooth solid curve shows the 6250\,\kms{}
broadened and scaled Fe\,{\sc ii} template superposed on the
continuum subtracted and He\,{\sc i} subtracted spectrum from the upper
panel. The lower of the two solid histograms is the residual
from subtracting this iron template fit. A simple (10,500\,\kms)
Gaussian fit to the He\,{\sc ii}\,$\lambda4686$ profile (dot-dashed)
and residuals with the Fe\,{\sc ii} template and He\,{\sc ii} fit
subtraction is also shown. Weak residuals remain at 4543\,\AA.}
\end{figure}

%
\begin{figure}
\plotfiddle{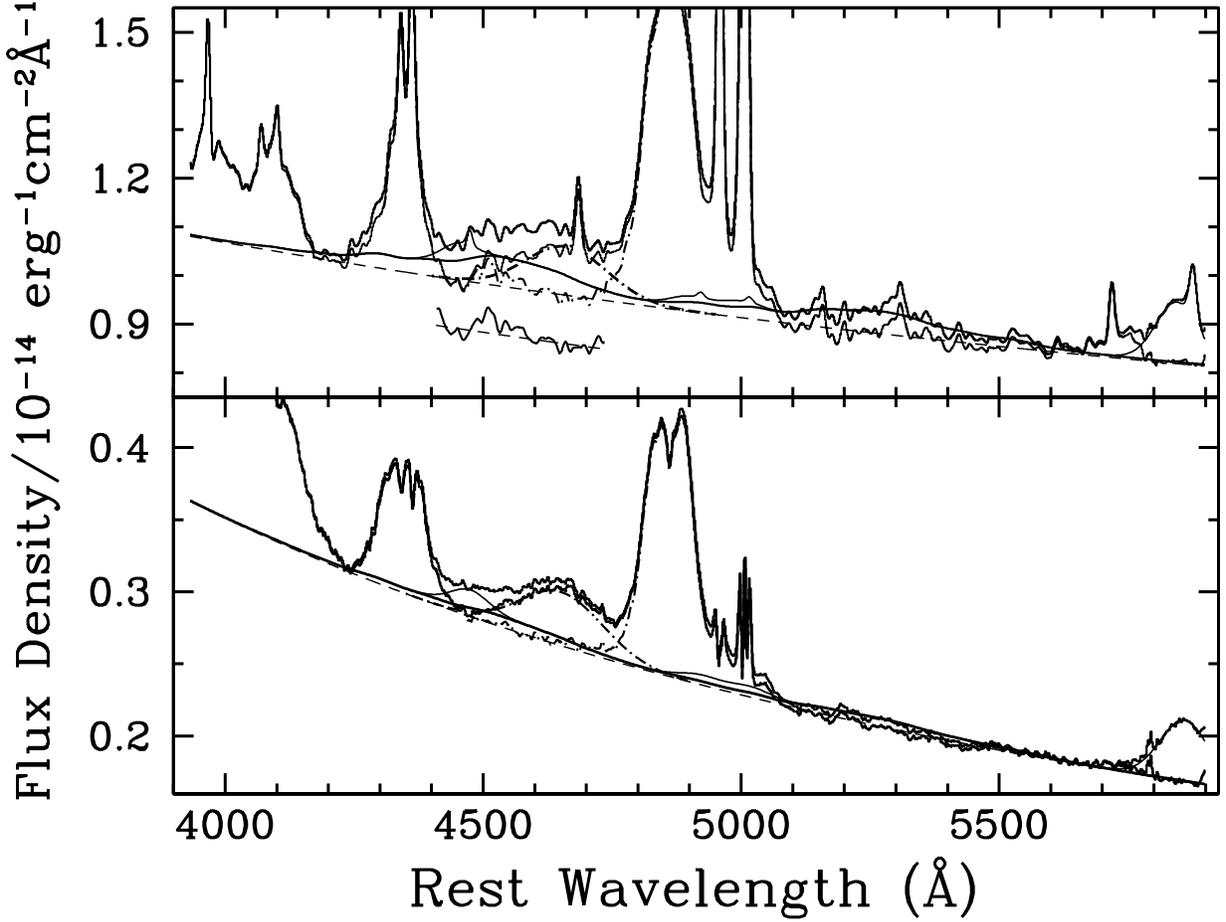}{50pt}{270}{358}{471}{0}{0}
\caption{The mean (upper panel) and rms (lower panel)
optical spectra of NGC 5548 in the rest frame.
In both cases, the smooth solid thick line shows the
broadened  (6250\,\kms{} FWHM) 
and scaled Fe\,{\sc ii} template, 
the dashed line shows a fit to the underlying continuum,
and the dash-dot line shows a Gaussian fit to the residual
He\,{\sc ii}\,$\lambda4686$ profile. 
As in Fig.\ 4, the thin
solid lines atop the Fe\,{\sc ii} template fit show the four
He\,{\sc i}\,$\lambda\lambda$\,4471, 4922, 5016, 5876 lines.
The spectrum after subtraction of the Fe\,{\sc ii} template 
and the He\,{\sc ii} fit is also shown. In the upper panel
a section of this residual is shown shifted downward 
for visibility.
The He\,{\sc ii} profile in the mean spectrum can be
fit with two Gaussian profiles of widths $\sim$800\,\kms\
and 12,000\,\kms, respectively. The width of the rms 
He\,{\sc ii} profile fit is 12,250\,\kms.}
\end{figure}

%
\begin{figure}
\plotfiddle{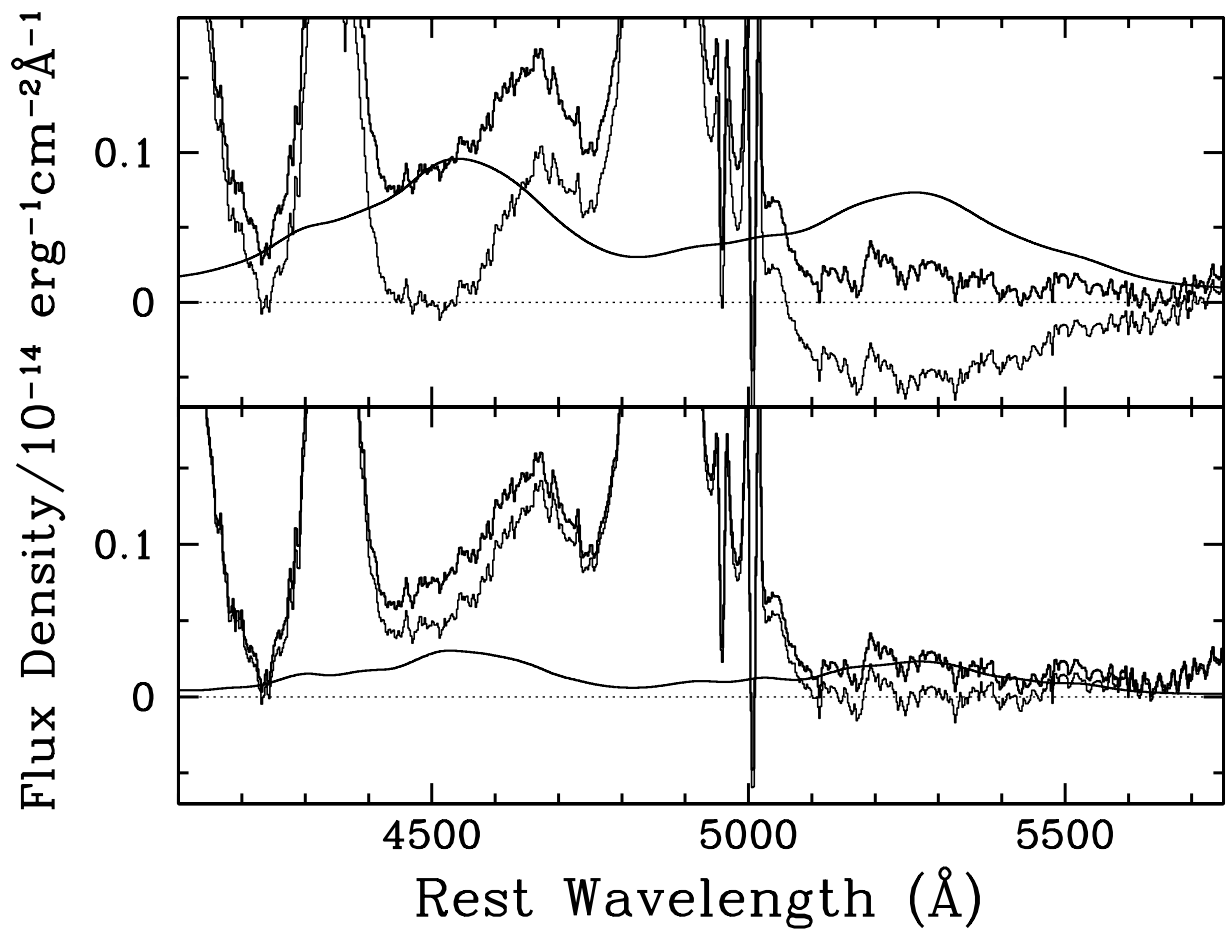}{50pt}{270}{358}{471}{0}{0}
\caption{
Sample fit with the V\'{e}ron-Cetty et al.\ (2004)
template to the difference spectrum shown in Fig.\ 5.  
In the upper panel, the template is fitted to match
the 4500\,\AA\ region only.  Notice the strong 
overprediction of the visible Fe\,{\sc ii}
features between 5050\,\AA\ and 5550\,\AA. 
The lower panel shows the template fit and residuals obtained
if the template is fit to the flux level of the $\sim$5300\,\AA\
Fe\,{\sc ii} transitions only.  This template poorly matches
the spectrum in the 4500\,\AA\ region.
The Fe\,{\sc ii} template has been broadened to
6250\,\kms.}
\end{figure}

%
\begin{figure}
\plotfiddle{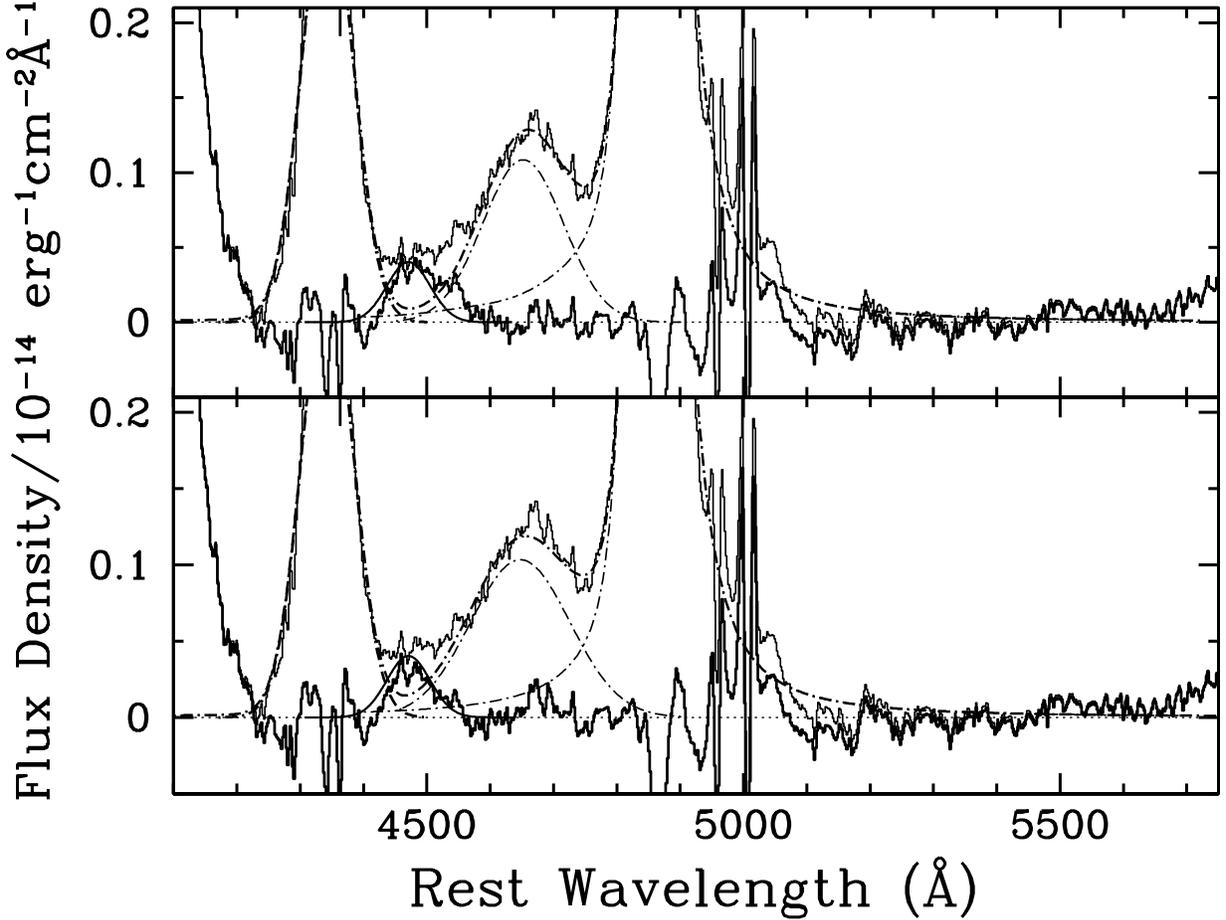}{50pt}{270}{358}{471}{0}{0}
\caption{
The continuum and Fe\,{\sc ii}-subtracted difference spectrum from the
lower panel of Fig.\ 7 is shown here with the residuals (thick line)
obtained when subtracting reasonable reproductions of the non-Fe
profiles (dot-dashed lines). The profile fits to \Hbeta{} and
\Hgamma{} were generated to reproduce especially the blue and red
sides of the core and the broad wings as allowed by the data; the
details of the fits to the line peaks can be ignored and the fit to
the [O\,{\sc iii}] residuals is suggestive only.  
The He\,{\sc ii}\,$\lambda4686$ profile is
shown with a 10,000\,\kms{} Gaussian profile (best fit) in the upper
panel, while the lower panel shows the widest He\,{\sc ii} profile of
12,000\,\kms{} allowed by the data. Neither fit can adequately
account for the excess emission at 4500\,\AA. 
However, reasonable consistency
is found with the expected strength of He\,{\sc i}\,$\lambda$4471
based on the observed He\,{\sc i}\,$\lambda$5876 feature. Here 
He\,{\sc i}\,$\lambda$4471 is represented by a Gaussian 
fit to He\,{\sc i}\,$\lambda$5876, shifted to 4471\AA\ (with fixed FWHM) and 
rescaled by a factor 0.3, a commonly established strength of 
He\,{\sc i}\,$\lambda$4471. This justifies our inclusion of the
He\,{\sc i} lines in our fitting of this region (e.g., Figs.\ 4, 5,
and 6).}
\end{figure}

%
\begin{figure}
\plotfiddle{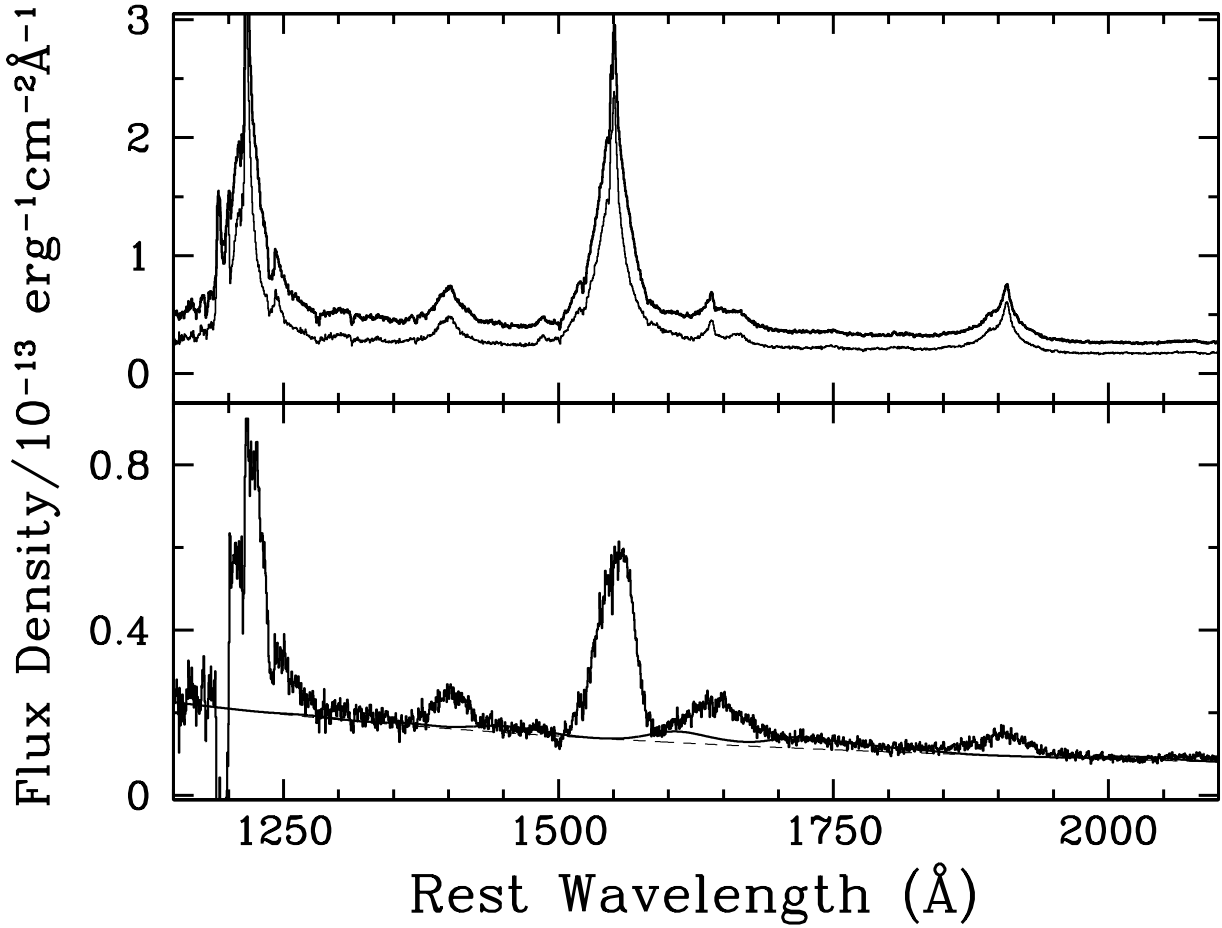}{50pt}{270}{358}{471}{0}{0}
\caption{The upper panel shows the high state and low state
ultraviolet spectra of NGC 5548 in the rest frame
as observed with {\em HST} in 1993. 
The lower panel shows the difference spectrum formed by
subtraction of the low state spectrum from the high
state spectrum. The smooth solid line shows the
broadened  (10,000\,\kms\ FWHM)
and scaled Fe\,{\sc ii} template, and 
the dashed line shows a fit to the underlying continuum.}
\end{figure}

%
\begin{figure}
\plotfiddle{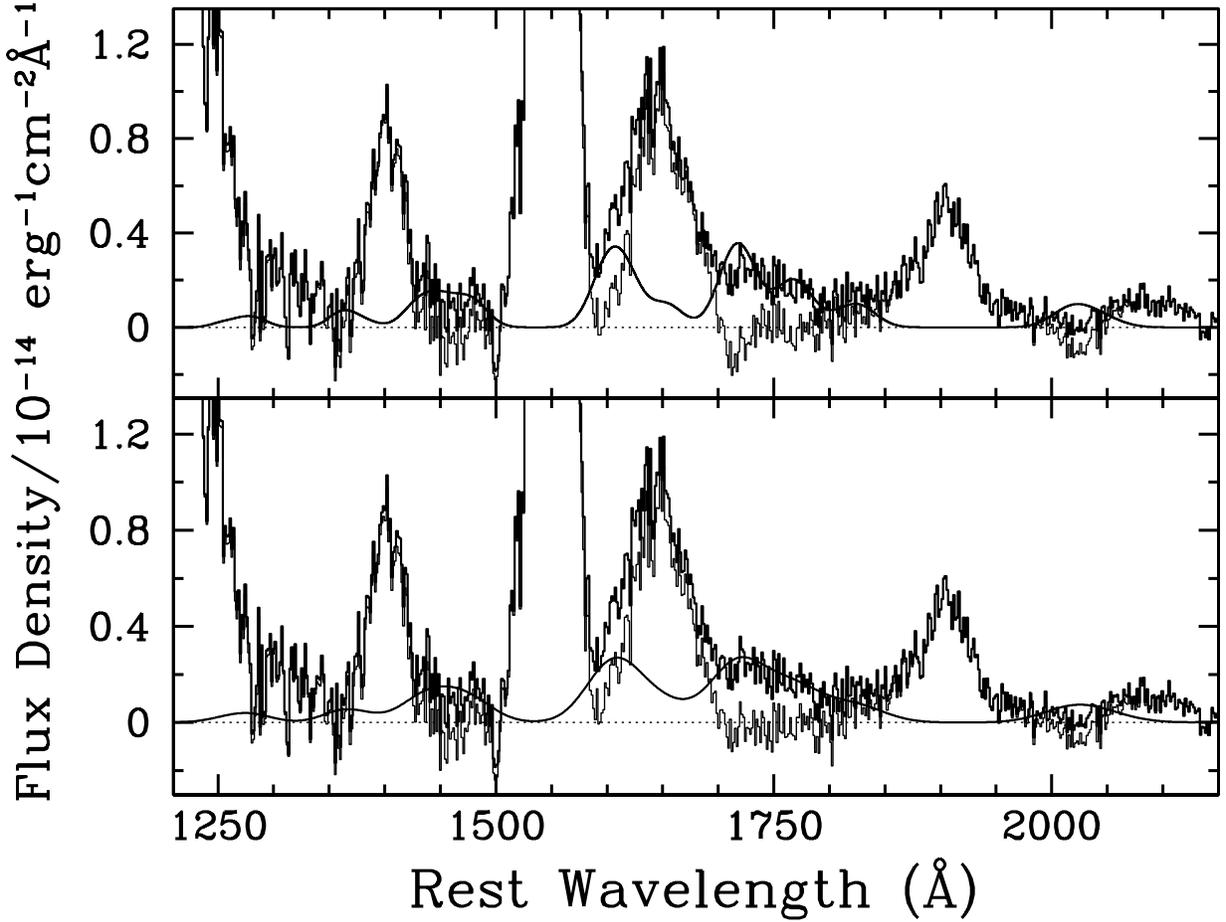}{50pt}{270}{358}{471}{0}{0}
\caption{An enlargement of the difference spectrum from the
lower panel of Fig.\ 9 with the continuum subtracted.
Again, the solid line shows the broadened and scaled Fe\,{\sc ii}
template, which has a width of 6250\,\kms\ in the upper panel
and of 10,000\,\kms\ in the lower panel. 
Note the poor fit of the 6250\,\kms\ iron template to the 1700\,\AA\
region. The template overpredicts the Fe\,{\sc ii} emission at 
2025\,\AA, possibly due to strong multiplets present in I\,Zw\,1
but not in NGC\,5548. The emission on either side of the expected
Fe\,{\sc ii} bump at 2025\,\AA\ is almost certainly Fe\,{\sc iii}
(see Vestergaard \& Wilkes 2001).}
\end{figure}

%
\begin{figure}
\plotfiddle{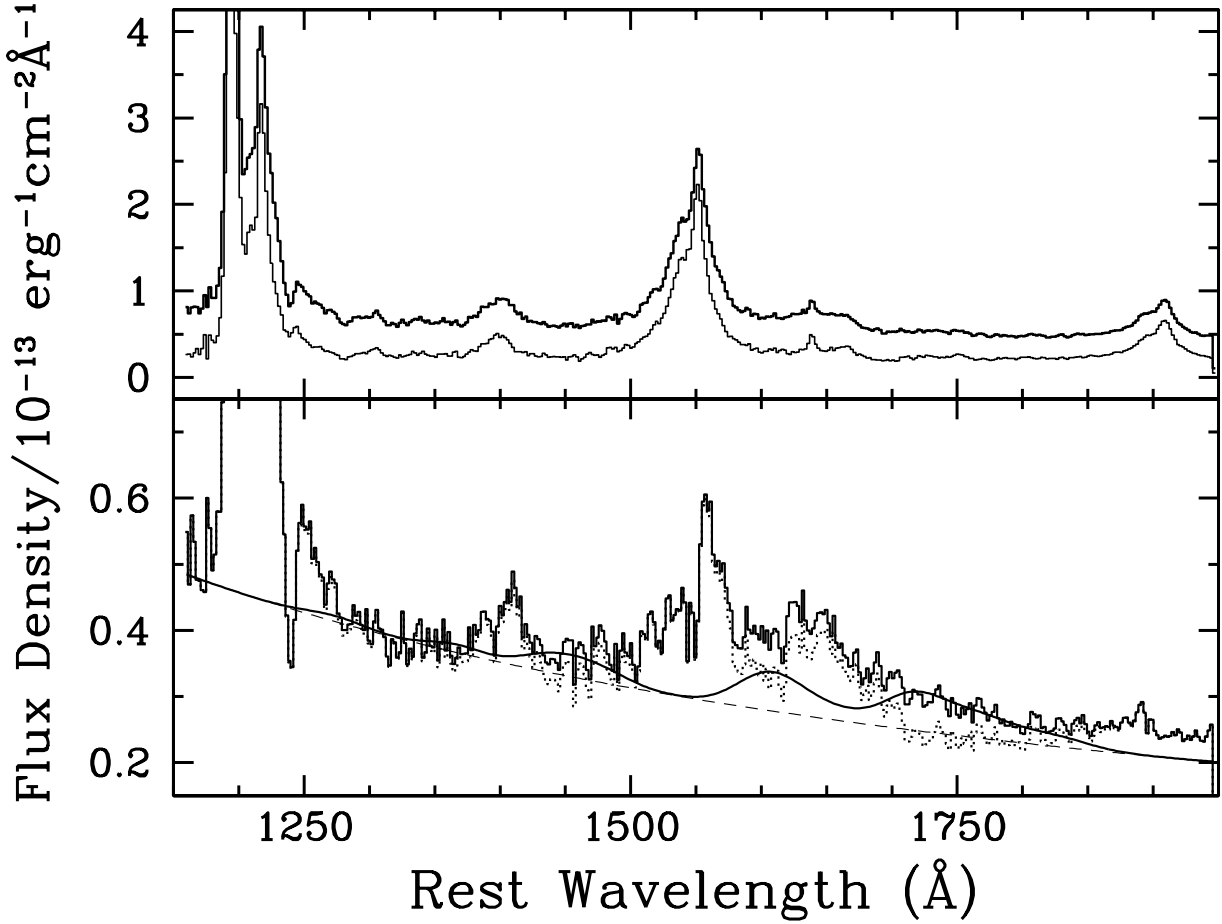}{50pt}{270}{358}{471}{0}{0}
\caption{The upper panel shows the high state and low state
ultraviolet spectra of NGC 5548 in the rest frame
as observed with {\em IUE} in 1988--89. 
The lower panel shows the difference spectrum formed by
subtraction of the low state spectrum from the high
state spectrum. The smooth solid line shows the
broadened (10,000\,\kms{} FWHM)
and scaled Fe\,{\sc ii} template, 
the dotted line represents the Fe\,{\sc ii} subtracted
spectrum,
and the dashed line shows a fit to the underlying continuum.}
\end{figure}

%
\begin{figure}
\plotfiddle{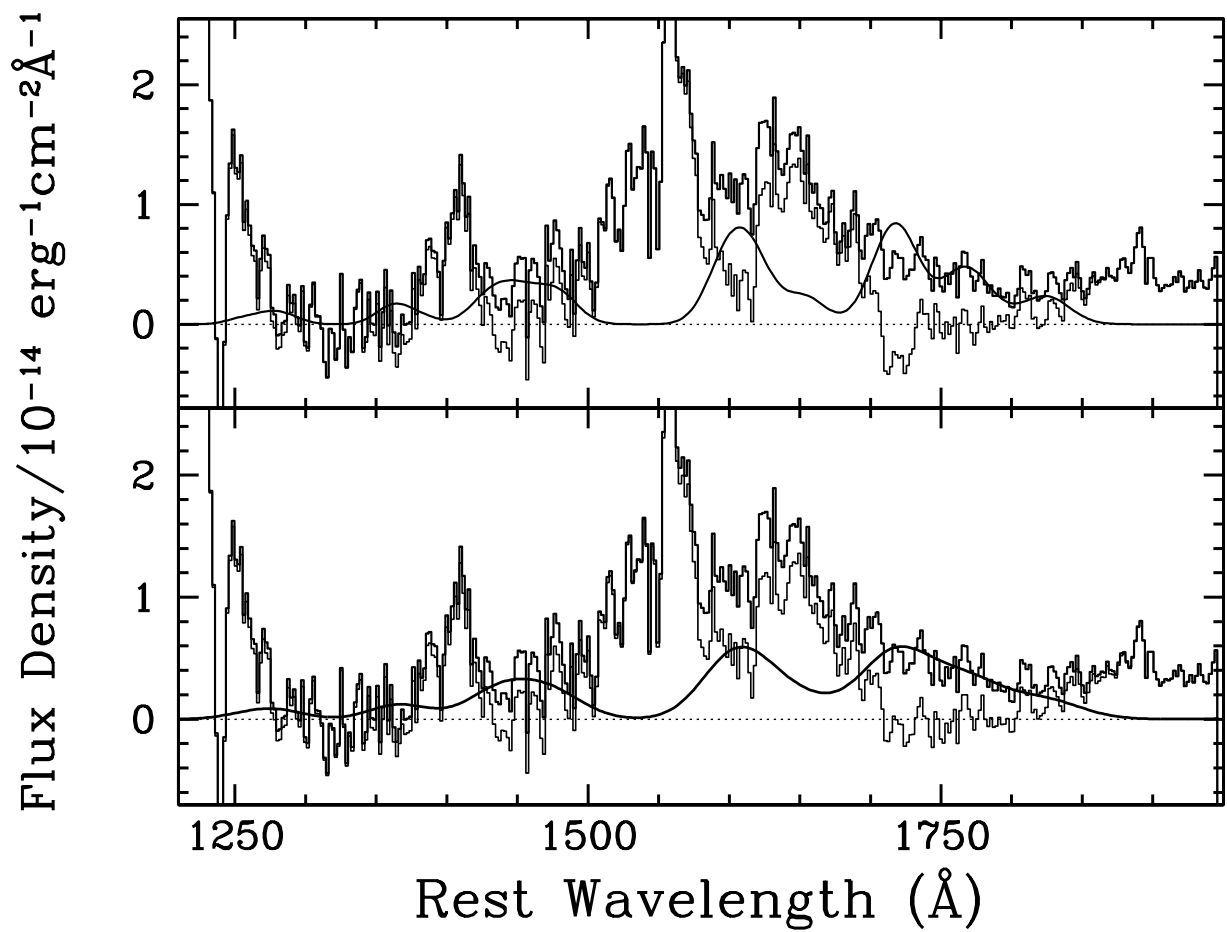}{50pt}{270}{358}{471}{0}{0}
\caption{An enlargement of the difference spectrum from the
lower panel of Fig.\ 11 with the continuum subtracted.
Again, the solid line shows the broadened and scaled Fe\,{\sc ii}
template, which has a width of 6250\,\kms\ in the upper panel
and of 10,000\,\kms\ in the lower panel.
As in Fig.\ 10, the template broadened to the rms width of
broad \Hbeta\ (6250\,\kms) yields a less satisfactory fit than 
that of the 10,000\,\kms\ template, especially at about 1700\,\AA.
The Fe\,{\sc ii} subtracted spectra are shown as the (lower) 
thin-lined histograms.}
\end{figure}

\clearpage

%
%
\begin{deluxetable}{cccc}
\tablewidth{0pt}
\tablecaption{Optical Fe\,{\sc ii}/H$\beta$ Flux Ratio}
\tablehead{
\colhead{Template Width} &
\colhead{Mean}  &
\colhead{Difference}  &
\colhead{RMS} \\
\colhead{(km s$^{-1}$)}    &
\colhead{Spectrum}    &
\colhead{Spectrum}    &
\colhead{Spectrum}    \\
\colhead{(1)} &
\colhead{(2)} &
\colhead{(3)} &
\colhead{(4)} 
}
\startdata
6250\tablenotemark{a} & 0.66 & 0.48 & 0.30 \\
\enddata
\tablenotetext{a}{Template fits with 10,000\,km s$^{-1}$ broadenings yield
very similar results.}
\end{deluxetable}

%
%
\begin{deluxetable}{ccccc}
\tablewidth{0pt}
\tablecaption{Ultraviolet Fe\,{\sc ii}/H$\beta$ Flux Ratio}
\tablehead{
\colhead{Template Width} &
\colhead{{\em HST} Mean}  &
\colhead{{\em HST} Difference}  &
\colhead{{\em IUE} Mean}  &
\colhead{{\em IUE} Difference} \\
\colhead{(km s$^{-1}$)}    &
\colhead{Spectrum}    &
\colhead{Spectrum}    &
\colhead{Spectrum}    &
\colhead{Spectrum}    \\
\colhead{(1)} &
\colhead{(2)} &
\colhead{(3)} &
\colhead{(4)} &
\colhead{(5)} 
}
\startdata
6250 & 0.45 & 3.12 & 0.56 & 4.98 \\
10000& 0.54 & 3.12 & 0.67 & 4.66 \\
\enddata
\end{deluxetable}

\end{document}